\documentclass[11pt]{article}
\usepackage{xcolor}
\pdfoutput=1 

\usepackage{jheppub} 

\usepackage[T1]{fontenc} 

\usepackage{amsmath,graphicx,xcolor,epsfig,slashed}
\usepackage{appendix}
\usepackage{tabularx}
\usepackage{float}
\usepackage{pdfpages}
\usepackage{epstopdf}
\usepackage{subfigure}
\usepackage{caption}
\usepackage{multirow}
\usepackage{todonotes}
\usepackage{graphicx}
\usepackage{subcaption}
\usepackage{ulem}

\title{Analysis of final state lepton polarization-dependent observables in $H\to \ell^{+}\ell^{-} \gamma$ in the SM at loop level}
\author[1]{Ishtiaq Ahmed,}
\author[2]{Usman Hasan,} 
\author[3]{Shahin Iqbal,}
\author[4]{M. Junaid,}
\author[5]{Bilal Tariq,} 
\author[6]{A. Uzair}
\affiliation{National Centre for Physics, Quaid-i-Azam University Campus, Islamabad.}
\date{September 2023}

\affiliation[1]{ishtiaq.ahmed@ncp.edu.pk}
\affiliation[2]{usman.hasan@ncp.edu.pk}
\affiliation[3]{shahin.iqbal@ncp.edu.pk}
\affiliation[4]{muhammad.junaid@ncp.edu.pk}
\affiliation[5]{bilal.tariq@ncp.edu.pk}
\affiliation[6]{ambreen.uzair@ncp.edu.pk}

\begin{document}

\abstract{
 Recently, the CMS and ATLAS collaborations have announced the results for $H\rightarrow Z[\rightarrow \ell^{+}\ell^{-}]\gamma$ with $\ell=e$ or $\mu$ \cite{CMS:2022ahq,CMS:2023mku},  where $H\rightarrow Z\gamma$ is a sub-process of $H\rightarrow \ell^{+} \ell^{-} \gamma$. This semi-leptonic Higgs decay receives loop induced resonant $H\rightarrow Z[\rightarrow \ell^{+}\ell^{-}]\gamma$ as well as non-resonant contributions. 
 To probe further features coming from these contributions to $H\rightarrow \ell^{+} \ell^{-} \gamma$, we argue that the polarization of the final state leptons is also an important parameter. We show that the contribution from the interference of resonant and non-resonant terms plays an important role when the polarization of final state lepton is taken into account, which is negligible in the case of unpolarized leptons. For this purpose, we have calculated the polarized decay rates and the longitudinal ($P_L$), normal ($P_N$) and transverse ($P_T$) polarization asymmetries. We find that these asymmetries purely come from the loop contributions and are helpful to further investigate the resonant and non-resonant nature of $H\rightarrow Z[\rightarrow \ell^{+}\ell^{-}]\gamma$ decay. We observe that for $\ell=e,\mu$, the longitudinal decay rate is highly suppressed around $m_{\ell\ell}\approx 60$GeV when the final lepton spin is $-\frac{1}{2}$, dramatically increasing the corresponding lepton polarization asymmetries. Furthermore, we analyze another observable, the ratio of decay rates $R^{\ell\ell'}_{i\pm}$, where $\ell$ and $\ell'$ refer to different final state lepton generations. Precise measurements of these observables at the HL-LHC and the planned $e^{+}e^{-}$ can provide a fertile ground to test not only the SM but also to examine the signatures of possible NP beyond the SM. 
}
\maketitle 
\section{Introduction}
The discovery of the Higgs particle in 2012 by the ATLAS \cite{ATLAS:2012yve} and CMS \cite{CMS:2012qbp} experiments was a great leap in the experimental validation of the the Glashow-Salam-Weinberg (GSW) theory of Weak interactions. GSW forms the $SU(2)\otimes U(1)$ part of the Standard Model (SM)  gauge group and combines the Weak and Electromagnetic interactions as different manifestations of a single, unified force, the Electroweak force. As it is currently formulated, the SM starts with assigning zero mass to all fundamental particles, which then get their masses through their interactions with the Higgs field (Higgs Mechanism) and particle masses are directly proportional to the sizes of their respective Higgs couplings. Therefore, using the measured masses of fundamental particles, the values of the Higgs couplings can be calculated within the SM with great precision.
With the discovery of the Higgs particle at 125 GeV, the next major goal in particle physics is to verify whether it is indeed the SM Higgs boson, or a new scalar particle as proposed in many models of particle physics that go beyond the SM. One way to answer this question is to study the many decays of the Higgs boson into various SM particles and calculate observable quantities, such as branching ratios, forward-backward asymmetries, final state polarization asymmetries, and so on. Indeed, one of the main goals of the future colliders such as the High Luminosity Large Hadron Collider (HL-LHC), and the International Linear Collider (ILC) is the precise measurement of the properties of the Higgs particle at 125GeV \cite{Liss:2013hbb,CMS:2013xfa}.  So far the Higgs couplings to $W$, $Z$, $\tau$, $b$, $t$ and photons have been measured experimentally by the CMS \cite{CMS:2017zyp, CMS:2018zzl, CMS:2019ekd, CMS:2013fjq, CMS:2018nsn, CMS:2018fdh, CMS:2018piu} and ATLAS \cite{ATLAS:2015muc, ATLAS:2014aga, ATLAS:2014xzb, ATLAS:2018pgp, ATLAS:2015xst, ATLAS:2018ynr, ATLAS:2018kot, ATLAS:2018mme, ATLAS:2018hxb} collaborations. It is worthwhile to mention here that the recently analyzed data \cite{CMS:2022ahq,CMS:2023mku} at a luminosity of $150 \text{ fb}^{-1}$ is  not sufficient to measure spin asymmetries. However, HL-LHC, which plans to increase the luminosity to $450-3000 \text{ fb}^{-1}$, as well as the planned $e^{+}e^{-}$ collider would allow better measurements of these asymmetries.  \cite{Schmidt:2016jra, Rossi:2019swj}

Among the various decay channels of the Higgs boson allowed within the SM, the semileptonic $H\rightarrow l^{+}l^{-}\gamma$ is particularly interesting.  Even though $H\rightarrow l^{+} l^{-}$ is a cleaner channel, however, it is suppressed by the Higgs coupling to the lepton, $y_{\ell}=\frac{m_{\ell}}{v}$. Here $m_{\ell}$ is the lepton mass and $v$ is the Higgs vacuum expectation value. The tree-level contribution for $H\rightarrow \ell^{+} \ell^{-}\gamma$ is also proportional to $y_{l}$ and suffers the same suppression for lighter leptons, but the semi-leptonic decay gets large contributions from the loop diagrams (running heavy SM particles $t$, $W$,$Z$), and the corresponding decay rates for lighter leptons are dominated by these loop contributions. Other than these one-loop contributions, when Higgs decays into a final state $Z$ boson and a photon, the subsequent leptonic decay $Z\rightarrow \ell^{+}\ell^{-}$ makes it a sub-process of $H\rightarrow \ell^{+}\ell^{-}\gamma$ i.e. $H\rightarrow Z[\rightarrow \ell^{+}\ell^{-}]\gamma$.  On the experimental side, $H\rightarrow \ell^{+}\ell^{-}\gamma$ has received significant attention over the years. Both ATLAS and CMS collaborations have performed searches of $H\rightarrow \ell^{+}\ell^{-}\gamma$ with $l=e,\mu$ for $\sqrt{s}=7,8$ and $13$ TeV \cite{ATLAS:2014fxe,CMS:2013rmy,CMS:2018myz, ATLAS:2020qcv} and again most recently by the CMS collaboration \cite{CMS:2022ahq}. 
 
 $H\rightarrow \ell^{+}\ell^{-}\gamma$ decay rates have been calculated in several previous studies \cite{Abbasabadi_1997,Chen_2013,Dicus_2013,Passarino:2013nka}. However, discrepancies still exist in these results \cite{Kachanovich:2020xyg}, motivating consideration of complementary observables such as polarization asymmetries. In section 3.2.1, we will discuss how the present work may help to ameliorate some of these discrepancies.

In total, up to one loop level, the radiative leptonic decay of Higgs gets contributions from (i) tree-level, (ii) one-loop box diagrams, (iii) those with an off-shell $Z$ boson, $H\rightarrow Z^{*}[\rightarrow \ell^{+}\ell^{-}]\gamma$, (iv) those with an off-shell photon  $H\rightarrow \gamma^{*}[\rightarrow \ell^{+}\ell^{-}]\gamma$ and finally (v) the `resonant' contribution from the on-shell $H\rightarrow Z[\rightarrow \ell^{+}\ell^{-}]\gamma$ \cite{Kachanovich:2021pvx}. In a previous exploratory study \cite{Akbar:2014pta}, the tree-level and some of the one-loop contributions were considered in the calculation of polarization asymmetry in $H\to \tau^{+}\tau^{-}\gamma$. However, in that case the decay rate is dominated by the tree-level contribution, and loop contributions are negligible. 

Here we present a \textit{complete} calculation of semileptonic decay rates \textit{including} all contributions up to one-loop level in the $R_{\xi}$ guage \cite{Kachanovich:2020xyg}, separating the resonant and non resonant contributions \cite{Kachanovich:2021pvx} and taking into account the polarization of final state leptons.  As we will see in section \ref{polrate} that once one considers the polarized decay rates, i.e. considering a specific polarization of the final state lepton, or otherwise looking at the final state lepton polarization asymmetry, very interesting and previously unknown behavior is revealed. Specifically, we find that the contribution to decay rates from loop diagrams shows a sharp suppression around $m_{\ell\ell}\approx 60$GeV for the case where \textit{final state lepton has longitudinal polarization $-\frac{1}{2}$}. On the other hand, we find that the tree-level contributions do not show any such suppression. Consequently, when considering the total longitudinal decay rates, this suppression is most pronounced for the case of final state electrons $\ell=e$, (due to the highly suppressed tree-level contribution), and nearly negligible for $\ell=\tau$ (where the tree-level dominates the total rates).


The structure of the paper is as follows: We start section \ref{theory} by presenting the theoretical framework required for $H\rightarrow \ell^{+}\ell^{-}\gamma$, in particular, calculations of the resonant and non-resonant contributions. In the section \ref{analysis}, we present details of our phenomenological analysis, including our calculations of the lepton polarization asymmetries and electron-to-muon ratios. Finally, in the section \ref{con} we present our conclusions.

\section{Theoretical frame work}\label{theory}
 In this section, we describe theoretical frame work to calculate the observables under consideration.
The tree level amplitude for the process, $H\to\ell^+\ell^-\gamma$, where $\ell$ corresponds to $e,\mu,\tau$, can be expressed as \cite{Sun:2013rqa},
\begin{eqnarray}
\mathcal{M}_{tree}&=&\mathcal{C}_{0}
\bar{u}(p_2)\bigg(\frac{2p_{2}^{\nu}+\gamma^{\nu}\slashed k}{2p_{2}\cdot k}-\frac{\slashed k \gamma^{\nu}+2p_{1}^{\nu}}{2p_{1} \cdot k}
\bigg)v(p_1)\epsilon_{\nu}^{*}\,, \label{MT}
\end{eqnarray}
where $p_1, p_2, k$ and $\epsilon^{*\nu}$ are the four-momenta of lepton, anti-lepton, photon and polarization of photon, respectively. Here, $2p_2\cdot k=u-m_{\ell}^2$ and $2p_1\cdot k=t-m_{\ell}^2$ where $s,t,u$ are the Mandelstam variables defined as $s=(p_1+p_2)^2$, $t=(p_1+k)^2$ and $u=(p_2+k)^2$ and coefficient $\mathcal{C}_{0}=-\frac{4\pi \alpha
 m_\ell}{m_{W} sin\theta_{W}}$.
The amplitude of one-loop contribution can be written as \cite{Kachanovich:2020xyg,Kachanovich:2021pvx,VanOn:2021myp},

\begin{eqnarray}
\mathcal{M}_{Loop}&=&\epsilon^{*\nu}(k_{\mu}p_{1\nu}-g_{\mu\nu}(k.p_1))\bar{u}(p_2)(\mathcal{C}_{1}\gamma^{\mu}+\mathcal{C}_{2}\gamma^{\mu}\gamma^{5})v(p_{1})\notag\\
&&+\epsilon^{*\nu}(k_{\mu}p_{2\nu}-g_{\mu\nu}(k.p_2))\bar{u}(p_2)(\mathcal{C}_{3}\gamma^{\mu}+\mathcal{C}_{4}\gamma^{\mu}\gamma^{5})v(p_1),\label{ML}
\end{eqnarray}
with
\begin{eqnarray}
\mathcal C_{1}&=&\frac{a_1+b_1}{2} \qquad \qquad
\mathcal C_{2}=\frac{a_1-b_1}{2},\notag \\
\mathcal C_{3}&=&\frac{a_2+b_2}{2} \qquad \qquad \mathcal C_{4}=\frac{a_2-b_2}{2}.\label{Cfunctions}
\end{eqnarray}
Here $a_i,b_i$ (where $i=1,2$) are functions of Mandelstam variables $s, t, u$ and are given in terms of the Passarino-Veltman decomposition of the tensor integrals \cite{Kachanovich:2021pvx} calculated in the limit of massless leptons. The  scalar one-loop form of these functions can be written in terms of Passerino-Veltman functions $B_{0}$ (with $\frac{1}{\epsilon}$ divergence), $C_0$ and $D_0$. By setting $D=4-2\epsilon$, the $1/\epsilon$ pole vanishes from the  $a_1$ and $b_1$ while expanding these to the order of $\epsilon^0$, these $a_{1,2}$ and $b_{1,2}$ expressions without UV-divergent pole are given in Appendix A.
 
 Using the amplitudes given in Eqs. \eqref{MT} and \eqref{ML}, we have derived the expressions for the amplitude square, $\vert\mathcal{M}\vert^2$, for the tree level, one-loop level and interference between tree and loop level contributions for the process $H \rightarrow \ell^{+}\ell^{-}\gamma$. These expressions are given as follows:
\begin{eqnarray}
\vert\mathcal{M}_{tree}\vert^2&=&16\mathcal{C}_0^2\bigg[\frac{\left(9 m_{\ell}^4+m_{\ell}^2 (-2 s+t-3 u)+t
   u\right)}{\left(m_{\ell}^2-t\right)^2}+\frac{\left(9 m_{\ell}^4+m_{\ell}^2 (-2 s-3 t+u)+t
   u\right)}{\left(m_{\ell}^2-u\right)^2}\notag\\
   &&+\frac{2 \left(17 m_{\ell}^4-8 m_{\ell}^2
   s-5 m_{\ell}^2 t-5 m_{\ell}^2 u+s^2+s
   t+s u+t u\right)}{\left(m_{\ell}^2-t\right)
   \left(m_{\ell}^2-u\right)}\bigg],
\end{eqnarray}
\begin{eqnarray}
\vert\mathcal{M}_{loop}\vert^2&=& s[t^2(\vert \mathcal{C}_1\vert^2+\vert \mathcal{C}_2\vert^2)+u^2(\vert \mathcal{C}_3\vert^2+\mathcal{C}_4\vert^2)]-2m_{\ell}^2[st\vert \mathcal{C}_1\vert^2+(st+2t^2)\vert \mathcal{C}_2\vert^2\notag \\
&&+su\vert \mathcal{C}_3\vert^2+(su+2u^2)\vert \mathcal{C}_4\vert^2-\text{Re}[\mathcal{C}_1\mathcal{C}_3^*](t+u)^2-\text{Re}[\mathcal{C}_2\mathcal{C}_4^*](t-u)^2]\notag \\
&&+m_{\ell}^4[s(\vert \mathcal{C}_1\vert^2+\vert \mathcal{C}_2\vert^2+\vert \mathcal{C}_3\vert^2+\vert \mathcal{C}_4\vert^2)+8t (\vert \mathcal{C}_2\vert^2+\vert \mathcal{C}_4\vert^2)-8(t+u)\text{Re}[\mathcal{C}_1\mathcal{C}_3^*]]\notag \\
&&-4m_{\ell}^6[ (\vert \mathcal{C}_2\vert^2+\vert \mathcal{C}_4\vert^2)-\text{Re}[\mathcal{C}_1\mathcal{C}_3^*]],\label{res5}
\end{eqnarray}
\begin{eqnarray}
\vert\mathcal{M}_{tree-loop}\vert^2&=&\frac{8 m_\ell }{\left(m_{\ell}^2-t\right) \left(m_{\ell}^2-u\right)}\bigg[[12 m_{\ell}^6-2 m_{\ell}^4 (s+8 t+6 u)+m_{\ell}^2 (t+u) (2 s+7t+3 u)\notag \\
&&-t \left(2 s u+(t+u)^2\right)]\text{Re}[\mathcal{C}_0\mathcal{C}_1^*]
+ [12 m_{\ell}^6-2 m_{\ell}^4 (s+6  t+8  u)\notag \\&&+ m_{\ell}^2(t+u)(2s+3t+7u))-u(2s t +(t+u)^2)]\text{Re}[\mathcal{C}_0\mathcal{C}_3^*]\bigg].
\end{eqnarray}
The double differential decay rate, over the variables $s$ and $t$, can be expressed as
\begin{eqnarray}
\frac{d^2\Gamma}{dsdt}=\frac{1}{256\pi^3m_{H}^3}\vert\mathcal{M}\vert^2,
\end{eqnarray}
where $\mathcal{M}=\mathcal{M}_{tree}+\mathcal{M}_{loop}$.

 In order to calculate the final state lepton polarization asymmetries, let us first introduce the orthogonal 4-vectors belonging to the
polarization of $\ell^-$  \cite{Kruger:1996cv,Fukae:1999ww}. These
polarization vectors, in their respective rest frames, can be defined as follow:
\begin{eqnarray}
S_L^{-}\equiv(0,\textbf{e}_L)&=&\left(0,\frac{\textbf{p}_1}{|\textbf{p}_1|}\right),\notag\\
S_N^{-}\equiv(0,\textbf{e}_N)&=&\left(0,\frac{\textbf{k}\times\textbf{p}_1}{|\textbf{k}\times\textbf{p}_1|}\right),\notag\\
S_T^{-}\equiv(0,\textbf{e}_T)&=&(0,\textbf{e}_N\times\textbf{e}_L),\label{PV1}
\end{eqnarray}
where the subscripts $L$, $N$ and $T$ correspond to the longitudinal, normal and transverse polarizations, respectively.
Also,
$\textbf{p}_1$, $\textbf{p}_2$ and $\textbf{k}$ denote the three momenta vectors of the
final particles $\ell^-$, $\ell^+$ and $\gamma$, respectively. The longitudinal unit vector
$S_{L}$ is boosted by Lorentz transformations in the direction of lepton momentum:
\begin{eqnarray}
 S_{L}^{-}&=&\left(\frac{|\textbf{p}_1|}{m_{\ell}},\frac{E_{l}\textbf{p}_1}{m_{\ell}|\textbf{p}_1|}\right).\label{PV3}
\end{eqnarray}

\subsection{Observables}\label{obs}
In the current study, we analyze: 

(i) the polarized decay rate, $\Gamma^{i\pm}$ which can be defined by the spin vectors given in Eqs. (\ref{PV1}) and (\ref{PV3}), where $i=L, N, T$, representing polarized decay rates in the longitudinal, normal and transverse directions, respectively. Furthermore, the double differential polarized decay rate 
can be written in the following form,
\begin{eqnarray}
\frac{d\Gamma^{i\pm}(s,t)}{dsdt} =\frac{1}{256\pi^3m_{H}^3}\vert\mathcal{M}^{i\pm}(s,t)\vert^2,\label{2.10}
\end{eqnarray}
 where $+(-)$ corresponds to the decay rate when final state lepton has $+\frac{1}{2}(-\frac{1}{2})$ polarization and $\mathcal{M}^{i\pm}(s,t)=\mathcal{M}^{i\pm}_{tree}(s,t)+\mathcal{M}^{i\pm}_{loop}(s,t)$. The expressions of  $\vert\mathcal{M}^{i\pm}(s,t)\vert^2$ are given in Appendix B in Eqs. (\ref{M21}-\ref{M26}).
 Moreover, the single differential decay rates can be defined as $d\Gamma^{i\pm}(s)\equiv\int\frac{d\Gamma^{i\pm}(s,t)}{dsdt}dt$ and $d\Gamma^{i\pm}(t)\equiv\int\frac{d\Gamma^{i\pm}(s,t)}{dsdt}ds$.
 
(ii) The lepton polarization asymmetry, $P_{i}$ which is related to the polarized differential decay rates is given in Eq. (\ref{2.10}) as follows:

\begin{eqnarray}
P_{i}(s) =\frac{d\Gamma^{i+}(s)-d\Gamma^{i-}(s)}{d\Gamma^{i+}(s)+d\Gamma^{i-}(s)},\qquad 
P_{i}(t)=\frac{d\Gamma^{i+}(t)-d\Gamma^{i-}(t)}{d\Gamma^{i+}(t)+d\Gamma^{i-}(t)}
\label{singlepol}
\end{eqnarray}

(iii) Using the definitions given in Eq. (\ref{2.10}), we define the lepton flavor universality (LFU) ratio $\mathcal{R}^{\ell\ell^\prime}_{i\pm}(s)$, as follows,
\begin{eqnarray}
\mathcal{R}^{\ell\ell^\prime}_{i\pm}(s)=\frac{d\Gamma^{\pm}(s)\vert_{H\to \ell^{+}\ell^{-}\gamma}}{d\Gamma^{\pm}(s)\vert_{H\to \ell^{'+}\ell^{'-}\gamma}}, \label{Ratio1}
\end{eqnarray}
which is the ratio between two decay rates when final leptons are different, where $\ell, \ell^\prime=e,\mu$ and $\tau$.

It is important to mention here that the square of the polarized amplitude at the tree level reads as  
\begin{eqnarray}
\vert\mathcal{M}^{i^\pm}_{tree}(s,t)\vert^2&=&\frac{\vert\mathcal{M}^{i}_{tree}(s,t)\vert^2}{2}\label{2.11}.
\end{eqnarray}
Consequently, the polarization asymmetry is absent at tree level. Moreover, the loop contributions $\vert\mathcal{M}^{i^\pm}_{loop}(s,t)\vert^2$ and the loop-tree interference contribution $\vert\mathcal{M}^{i\pm}_{tree-loop}\vert^2$ for $i=L,N,T$ are explicitly expressed in appendix.

\subsection{Resonant and non-resonant contributions}
We use the scheme for separating the resonant and non-resonant contributions to $H\rightarrow \ell^{+}\ell^{-}\gamma$, in our calculations, as discussed in Ref. \cite{Kachanovich:2021pvx}.
This goal can be achieved by separating the $\mathcal{C}_i$ given in Eq. (\ref{Cfunctions}) in terms of resonant and non-resonant parts as follows:
\begin{eqnarray}
    \mathcal{C}_i=\mathcal{C}_i^{res}+\mathcal{C}_i^{nr}\qquad i=1,\cdots,4,\label{CResNR}
\end{eqnarray}
where 
\begin{eqnarray}
    \mathcal{C}_i^{res}(s)\equiv\frac{\alpha(m_Z^2)+r\beta(m_Z^2)}{s-m_Z^2+im_Z\Gamma_Z};\label{res}
\end{eqnarray}
and $r=+(-)$ for $i=1,3(2,4)$. Whereas, $\alpha(s)$ and $\beta(s)$ are given in Ref. \cite{Kachanovich:2021pvx}. For the non-resonant part:

\begin{eqnarray}
   \mathcal{C}_i^{nr}(s,t)&\equiv& a_i'(s,t)+f[\alpha]+rf[\beta], \label{nres}
   \end{eqnarray}
   with
   \begin{eqnarray}
   f[\alpha]&=&\frac{\alpha(s)-\alpha(m_Z^2)}{s-m_Z^2+im_Z\Gamma_Z},\qquad f[\beta]=f[\alpha\rightarrow\beta],\qquad
   a_i'(s,t)=\Tilde{a}_i+r\Tilde{b}_i.
\end{eqnarray}

Here $\Tilde{a}_i$ and $\Tilde{b}_i$ are given in Ref. \cite{Kachanovich:2021pvx}. Therefore, the resonance contribution for unpolarized decay rate is,
\begin{eqnarray}
\frac{d\Gamma_{res}}{dsdt}&=&\frac{1}{256\pi^3m_{H}^3}\vert\mathcal{M}_{res}\vert^2,
\end{eqnarray} 
and for polarized decay rate is,
\begin{eqnarray}
    \frac{d\Gamma_{res}^{i\pm}}{dsdt}&=&\frac{1}{256\pi^3m_{H}^3}\vert\mathcal{M}_{res}^{i\pm}\vert^2 .\label{Res1}
\end{eqnarray}
 One can get the $\vert\mathcal{M}_{res}\vert^2$ and $\vert\mathcal{M}_{res}^{i\pm}\vert^2$ by replacing $\mathcal{C}_i$ with $\mathcal{C}_i^{res}$ from Eq. \eqref{res} in Eqs. (\ref{res5}, \ref{M21}, \ref{M23}, \ref{M25}). Similarly, the non resonance contributions to the unpolarized and polarized decay rates are,
\begin{eqnarray}
\frac{d\Gamma_{nr}}{dsdt}&=&\frac{1}{256\pi^3m_{H}^3}\vert\mathcal{M}_{nr}\vert^2,\qquad\qquad
 \frac{d\Gamma_{nr}^{i\pm}}{dsdt}=\frac{1}{256\pi^3m_{H}^3}\vert\mathcal{M}_{nr}^{i\pm}\vert^2,\label{Res2}
\end{eqnarray}
where $\vert\mathcal{M}_{nr}\vert^2$ and $\vert\mathcal{M}_{nr}^{i\pm}\vert^2$ can be found by replacing $\mathcal{C}_i$ with $\mathcal{C}_i^{nr}$ from Eq. \eqref{nres} in Eqs. (\ref{res5}, \ref{M21}, \ref{M23}, \ref{M25}).

\begin{table}[ht]
\begin{tabular}{ccc}
\hline\hline
$\alpha^{-1}=132.184$, $m_{H}=125.1$ GeV, & $m_W=80.379$ GeV, & $m_Z=91.1876$ GeV,\\
$m_{t}=173.1$ GeV, & $m_{\mu}=0.106$ GeV, & $m_e=0.51\times10^{-3}$ GeV,\\ $\Gamma_Z=2.4952$ GeV,  & $C_W=0.881469$, & $G_F=1.1663787\times10^{-5}$ GeV$^{2}$.\\
\hline\hline
\end{tabular}\label{input}
\centering \caption{Default values of input parameters used in the
calculations}
\label{Numericals}
\end{table}

\section{Phenomenological analysis}\label{analysis}
This section presents the analysis of observables under consideration. To get the numerical results of observables, we set the following limits for the Mandelstam variables $s$ and $t$:
\begin{eqnarray}
    s_{min}&=&4m_\ell^2,\qquad\qquad s_{max}=m_H^2,\notag \\
    t_{min(max)}&=&\frac{1}{2}\left(m_H^2-s+2m_\ell^2\mp(m_H^2-s)\sqrt{1-4m_\ell^2/s}\right).
    \end{eqnarray}
 The numerical values of various parameters are given in Table \ref{Numericals}. In Table \ref{Rate} we calculate the total decay rates by using following the kinematic cuts  \cite{Kachanovich:2021pvx,Kachanovich:2020xyg}:
 \begin{eqnarray}
    E_{\gamma , min}&=&5\text{GeV},\qquad\qquad s,t,u>(0.1m_H)^2.
    \end{eqnarray}
We have retained the finite leptons masses in the phase space integration.
\begin{table}[ht]
\begin{tabular}{c|c|c|c}
\hline\hline
$\Gamma_{\text{total}}\quad\&\quad\mathcal{B}r$ & $H\to e^+e^-\gamma$ & $H\to \mu^+\mu^-\gamma$ & $H\to \tau^+\tau^-\gamma$ \\
\hline 
$\Gamma_{\text{total}}(\text{GeV})$ & $2.426\times 10^{-7}$  & $2.685\times 10^{-7}$  & $7.448\times 10^{-6}$ \\
$\mathcal{B}r_{\text{total}}$ & $5.92\times 10^{-5}$  & $6.55\times 10^{-5}$  & $1.82\times 10^{-3}$ \\
\hline
$\Gamma^{L-}_{\text{total}}(\text{GeV})$ & $1.299\times 10^{-7}$  & $1.428\times 10^{-7}$  & $3.701\times 10^{-6}$ \\
$\mathcal{B}r^{L-}_{\text{total}}$ & $3.17\times 10^{-5}$  & $3.48\times 10^{-5}$  & $9.02\times 10^{-4}$ \\
\hline
$\Gamma^{L+}_{\text{total}}(\text{GeV})$ & $1.1266\times 10^{-7}$  & $1.2574\times 10^{-7}$  & $3.7467\times 10^{-6}$ \\
$\mathcal{B}r^{L+}_{\text{total}}$ & $2.75\times 10^{-5}$  & $3.07\times 10^{-5}$  & $9.14\times 10^{-4}$ \\
\hline
$\Gamma^{N-}_{\text{total}}(\text{GeV})$ & $1.2127\times 10^{-7}$  & $1.3392\times 10^{-7}$  & $3.7212\times 10^{-6}$ \\
$\mathcal{B}r^{N-}_{\text{total}}$ & $2.96\times 10^{-5}$  & $3.27\times 10^{-5}$  & $9.08\times 10^{-4}$ \\
\hline
$\Gamma^{N+}_{\text{total}}(\text{GeV})$ & $1.2128\times 10^{-7}$  & $1.3477\times 10^{-7}$  & $3.7270\times 10^{-6}$ \\
$\mathcal{B}r^{N+}_{\text{total}}$ & $2.96\times 10^{-5}$  & $3.29\times 10^{-5}$  & $9.09\times 10^{-4}$ \\
\hline
$\Gamma^{T-}_{\text{total}}(\text{GeV})$ & $1.2129\times 10^{-7}$  & $1.3779\times 10^{-7}$  & $3.8232\times 10^{-6}$ \\
$\mathcal{B}r^{T-}_{\text{total}}$ & $2.96\times 10^{-5}$  & $3.36\times 10^{-5}$  & $9.32\times 10^{-4}$ \\
\hline
$\Gamma^{T+}_{\text{total}}(\text{GeV})$ & $1.2126\times 10^{-7}$  & $1.3072\times 10^{-7}$  & $3.7180\times 10^{-6}$ \\
$\mathcal{B}r^{T+}_{\text{total}}$ & $2.96\times 10^{-5}$  & $3.19\times 10^{-5}$  & $9.07\times 10^{-4}$ \\
\hline\hline
\end{tabular}\label{input}
\centering \caption{Total decay rates and corresponding branching ratios of the processes under consideration. The uncertainties due to different parameters are less than $1\%$.}
\label{Rate}
\end{table}

\subsection{Plot scheme for polarized differential decay rates $d\Gamma^{i\pm}$}\label{polrate}

We present our results for the polarized differential decay rates $d\Gamma^{\pm}$ for the process $H\rightarrow \ell^+\ell^-\gamma$ against the di-lepton invariant mass $\sqrt{s}\equiv m_{\ell\ell}$  and the lepton-photon invariant mass $\sqrt{t}\equiv m_{\ell\gamma}$. Before discussing our results, first we present our plotting scheme: The solid lines show unpolarized differential decay rates, $d\Gamma$, while the polarized decay rates, $d\Gamma^{\pm}$ for the final state lepton with polarization $+\frac{1}{2}$ and $-\frac{1}{2}$, are represented by dotted  and dashed lines, respectively. The unpolarized total decay rate can be obtained by adding up the polarized decay rates $d\Gamma^{\pm}$, i.e., $d\Gamma_{tot}=d\Gamma^{+}+d\Gamma^{-}$. We find that our results for $d\Gamma_{tot}$ agree with the results presented in \cite{Kachanovich:2021pvx}. In the present work, we represent $d\Gamma_{tot}$ with a solid black line, while the tree-level contribution $d\Gamma_{tree}$ and the loop contribution $d\Gamma_{loop}$ are  represented by gray and red lines, respectively. The blue (orange) line represents resonant (non-resonant) contribution to the decay rate, $d\Gamma_{res(nr)}$, whereas, the green line shows the interference between resonance and non-resonance contributions $d\Gamma_{res-nr}$, while the magenta (brown) represents interference of tree and resonance (non-resonance) contribution, $d\Gamma_{tree-res(tree-nr)}$.

Note that at the tree-level, the polarized differential decay rate is equal to half of the unpolarized rate  as given in Eq. (\ref{2.11}). We have therefore chosen not to show the tree level contributions to polarized decay rates separately.

\subsubsection{Longitudinally polarized differential decay rates}

\begin{figure}
\centering\scalebox{1}{
\begin{tabular}{cc}
\includegraphics[width=3in,height=2in]{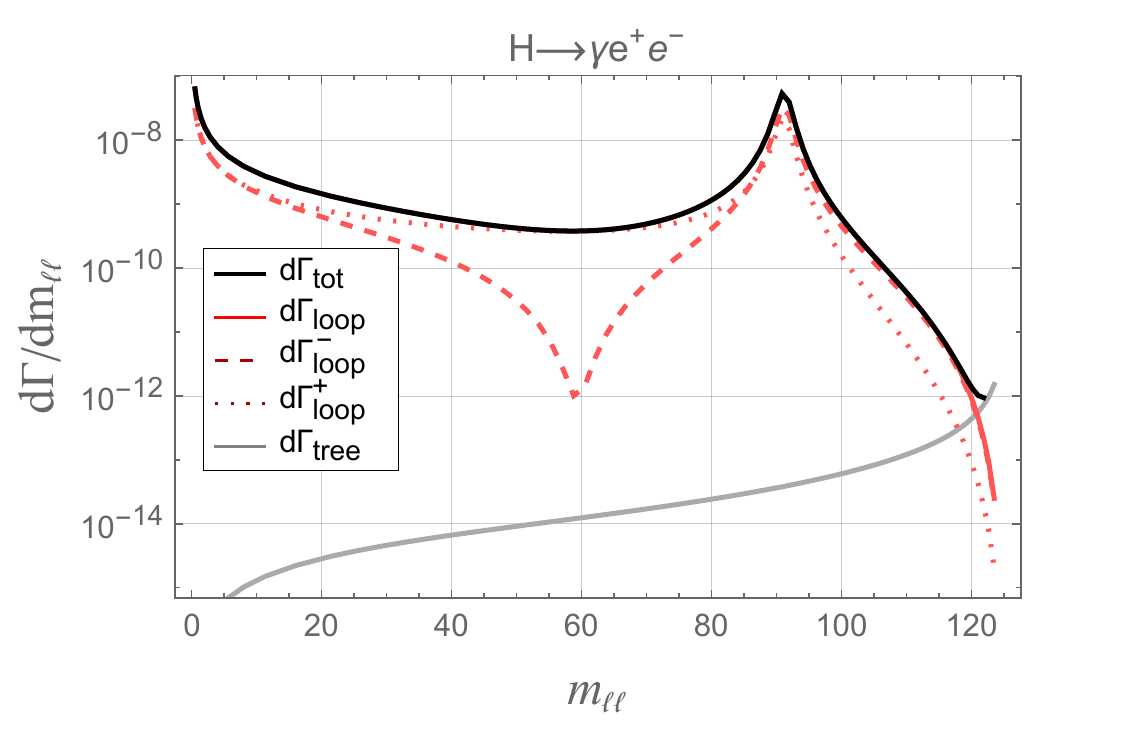} &
\includegraphics[width=3in,height=2in]{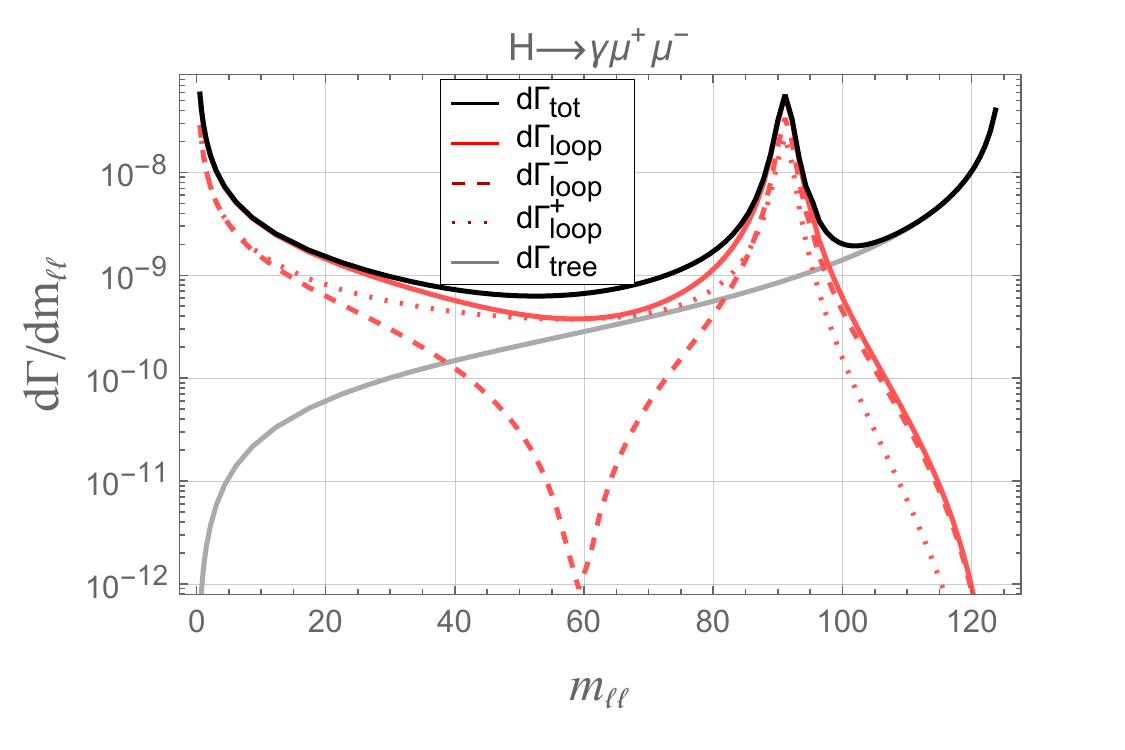}\\
(a) & (d) \\
\includegraphics[width=3in,height=2in]{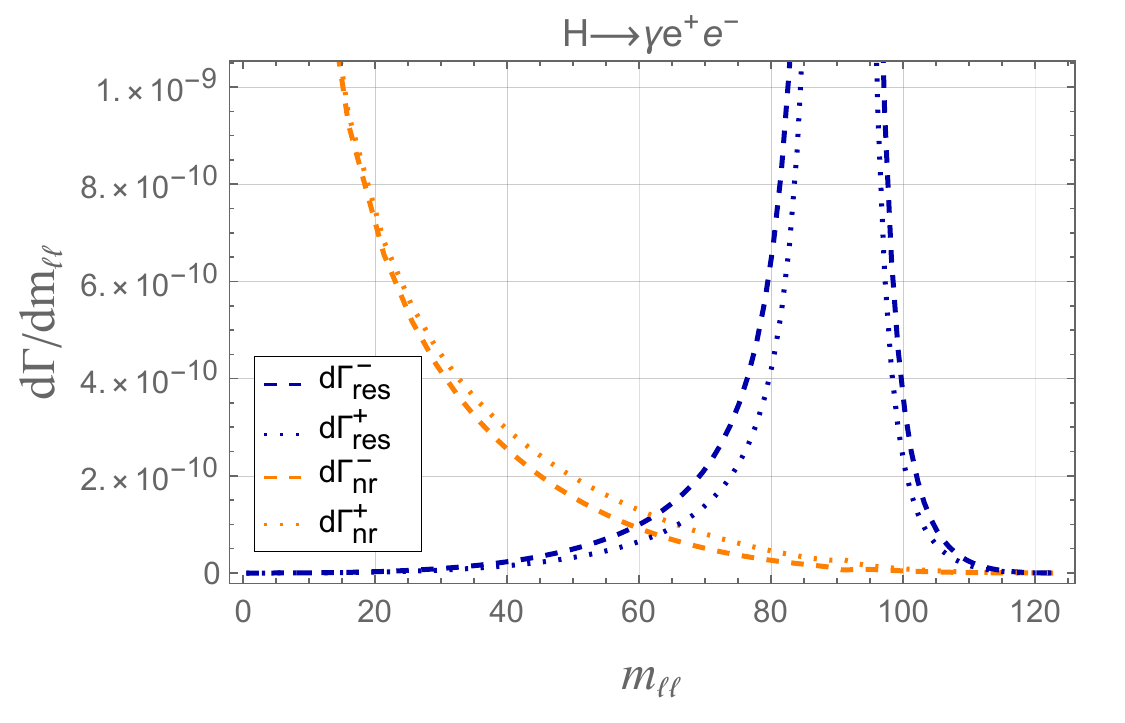} &
\includegraphics[width=3in,height=2in]{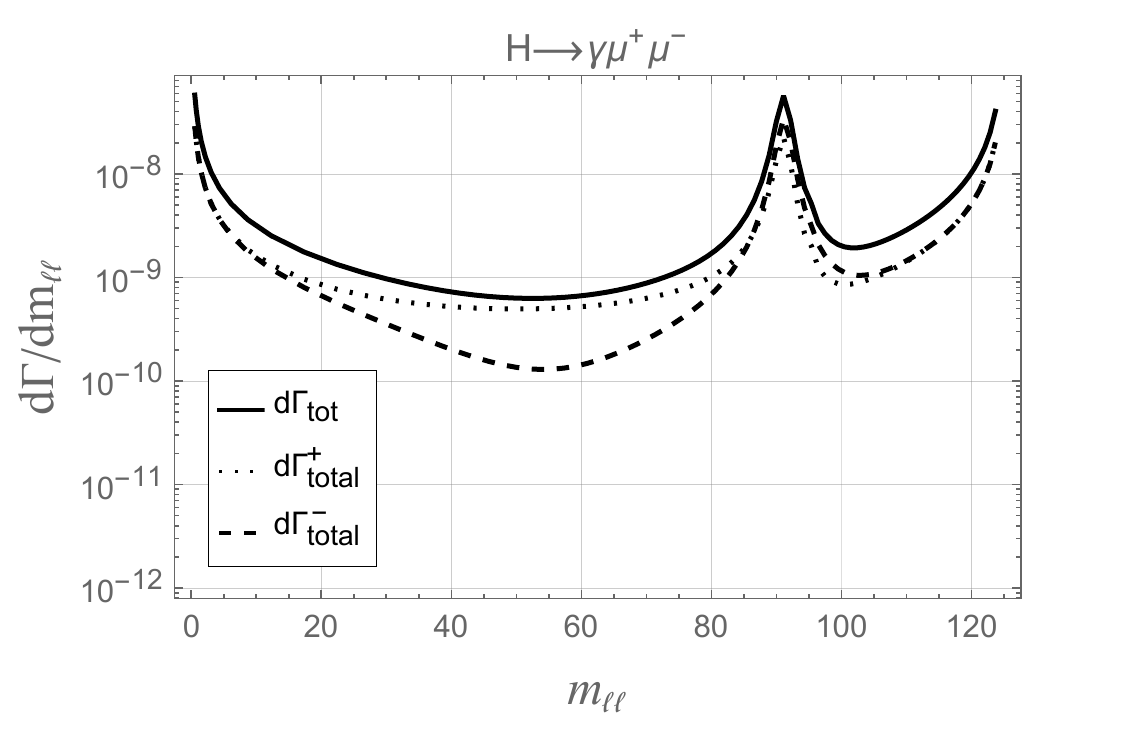} \\
(b) & (e) \\
\includegraphics[width=3in,height=2in]{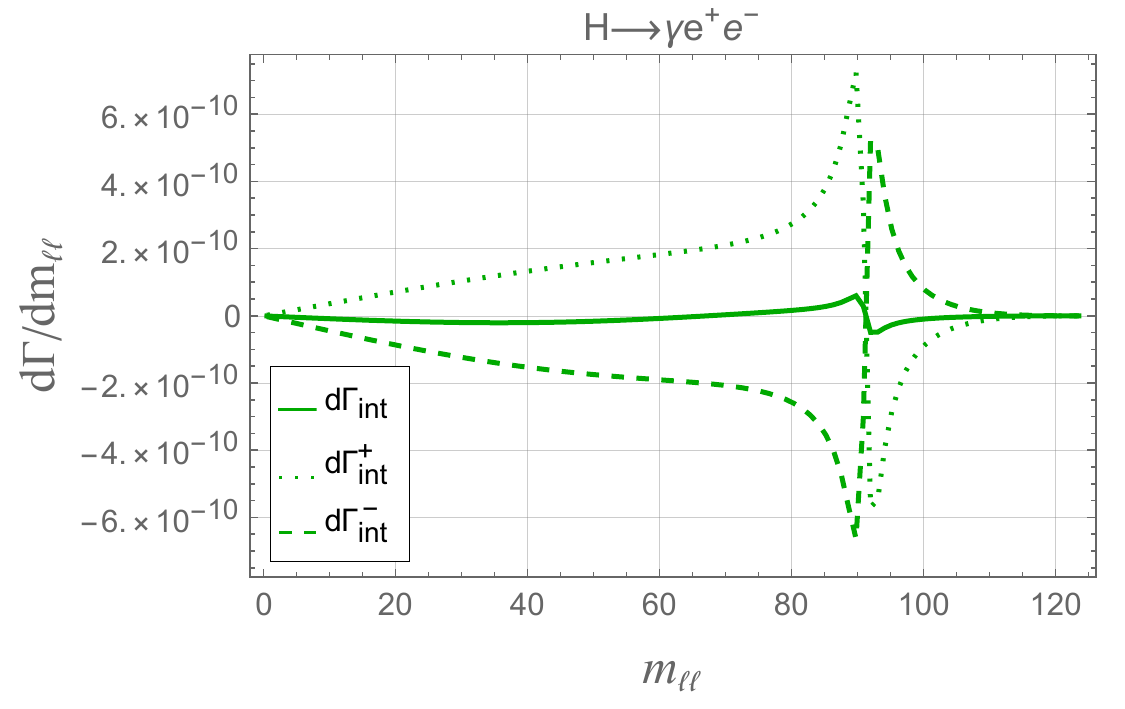} &
\includegraphics[width=3in,height=2in]{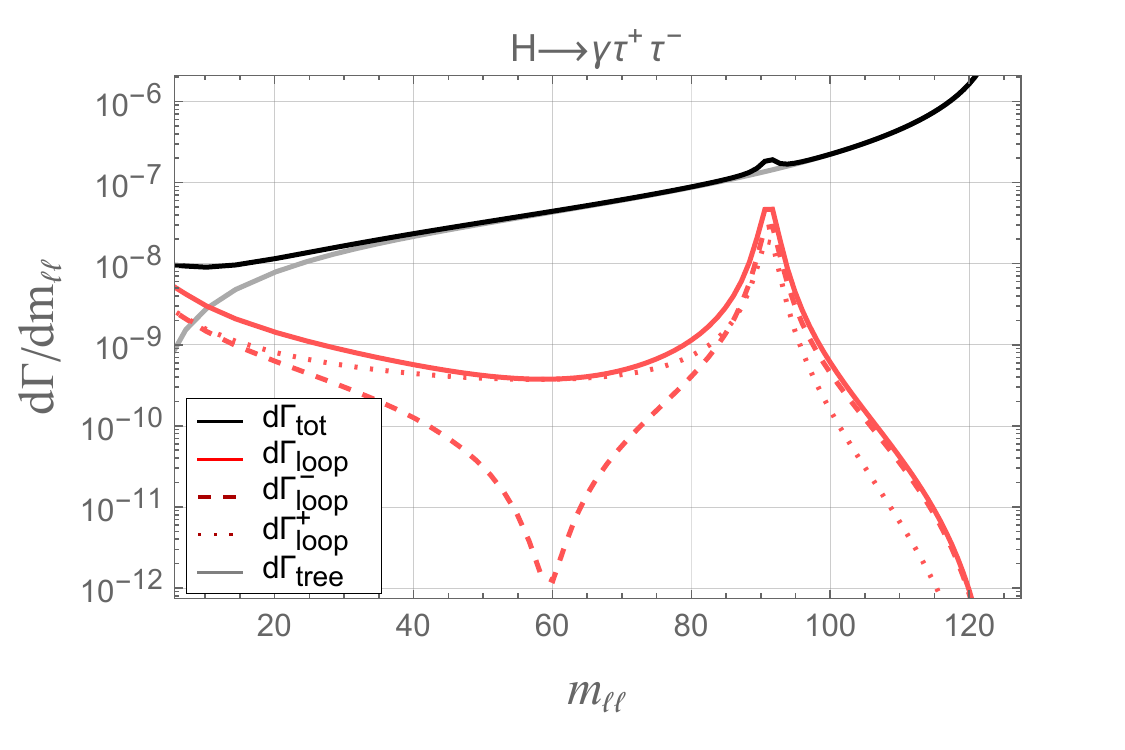}\\
(c) & (f)
\end{tabular}}
\caption{Longitudinally polarized differential decay rate with respect to invariant dilepton mass $m_{\ell\ell}$. (a,d,f) show  unpolarized total, loop and tree contributions for $\ell=e,\mu,\tau$ respectively. Polarized parts of loop are also shown. (b) shows polarized resonance and non-resonance terms while (c) shows the polarized and unpolarized interference term for $\ell=e$. (e) shows the polarized as well as unpolarized total decay rate for $\ell=\mu$. \textbf{Color scheme:} The solid lines show $d\Gamma$ while $d\Gamma^{\pm}$ for the final state lepton with polarization $+\frac{1}{2}$ and $-\frac{1}{2}$ are represented by dotted and dashed lines respectively. $d\Gamma_{tot}$, $d\Gamma_{tree}$ and $d\Gamma_{loop}$ are  represented by the black, gray and red lines respectively. The blue (orange) line represents $d\Gamma_{res(nr)}$ whereas the green line shows $d\Gamma_{res-nr}$.}
\label{LES}
\end{figure}

Fig. \ref{LES}(a) shows the decay rates $d\Gamma(m_{ee})$ for longitudinally polarized electrons in final state. As the tree-level contribution is negligible, therefore, the total polarized differential decay rate primarily receives contribution from the loop diagrams, i.e., $d\Gamma_{tot}\sim d\Gamma_{loop}$. In these figures one can notice that the loop contribution continues to dominate the total rate through out the kinematic region except at the end of the spectrum (near the Higgs pole), where the tree-level contribution increases and starts to dominate.
Note that the region between the photon and the $Z$ pole is suitable not only to test the SM but also probe the possible NP . In this region, we find a sharp dip in the decay rate $d\Gamma_{loop}^{L-}$ around $m_{ee}\sim 60$ GeV, however, the decay rate $d\Gamma_{loop}^{L+}$ shows smooth behavior. The reason of this dip is the interplay of resonance and non-resonance terms in the loop contribution.

The resonance $d\Gamma_{res}^{\pm}$, non-resonance $d\Gamma_{nr}^{\pm}$ and the interference of resonance and non-resonance $d\Gamma_{res-nr}^{\pm}$ contributions are shown separately in Figs. \ref{LES}(b,c). As can be seen in Fig. \ref{LES}(c), the interference contribution (solid green line) to unpolarized decay rate is negligible i.e. $d\Gamma_{res-nr}\ll d\Gamma_{tot}$ which is consistent with the results given in \cite{Kachanovich:2021pvx}. 
However, the interference contributions $d\Gamma_{res-nr}^{\pm}$ to the polarized decay rates is comparable in size (but with opposite sign for $d\Gamma_{res-nr}^{-}$) to the resonant $d\Gamma_{res}^{\pm}$, and non-resonant  $d\Gamma_{nr}^{\pm}$ contributions, as shown in fig.  \ref{LES}(c). One can write,


 \begin{eqnarray}
    \frac{d\Gamma^{\pm}_{loop}}{ds}= \frac{d\Gamma^{\pm}_{res}}{ds}+ \frac{d\Gamma^{\pm}_{nr}}{ds}+ \frac{d\Gamma^{\pm}_{res-nr}}{ds}.
\end{eqnarray}
 
Around $m_{ee}\approx 60$ GeV region, the negative contribution $d\Gamma_{res-nr}^{-}$ nearly cancels the positive contribution of $d\Gamma_{res}^{-}+d\Gamma_{nr}^{-}$. Consequently, the polarized decay rate $d\Gamma_{loop}^{-}$ shows a dip as shown in Fig. \ref{LES}(a). However, the contribution $d\Gamma_{res-nr}^{+}$ remains positive and thus does not produce any cancellation. 

\begin{figure}
\centering
\begin{tabular}{ccc}
\includegraphics[width=3in,height=2in]{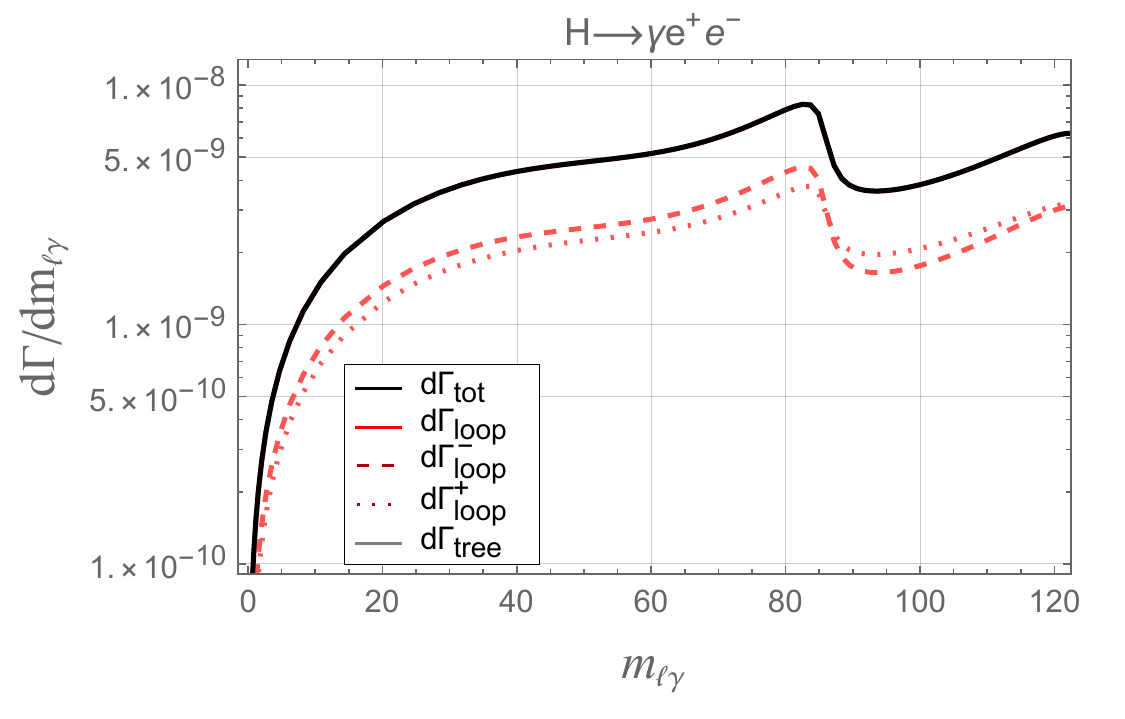}&
\includegraphics[width=3in,height=2in]{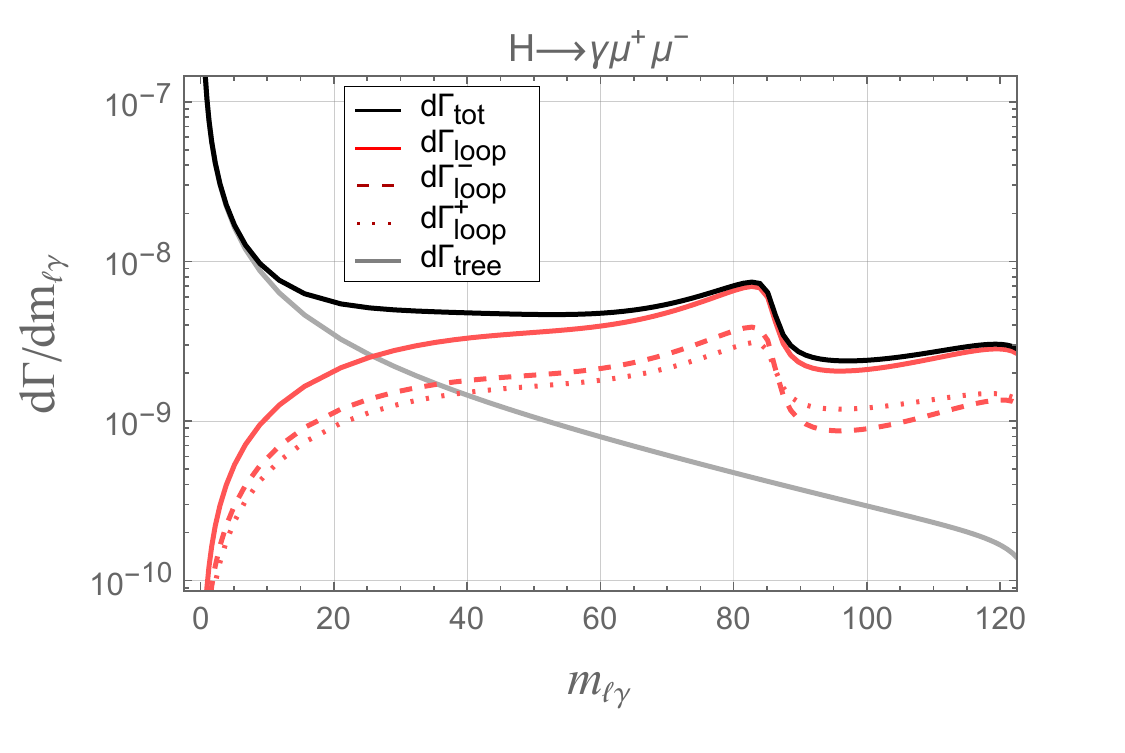}&\\
(a)&(d) \\
\includegraphics[width=3in,height=2in]{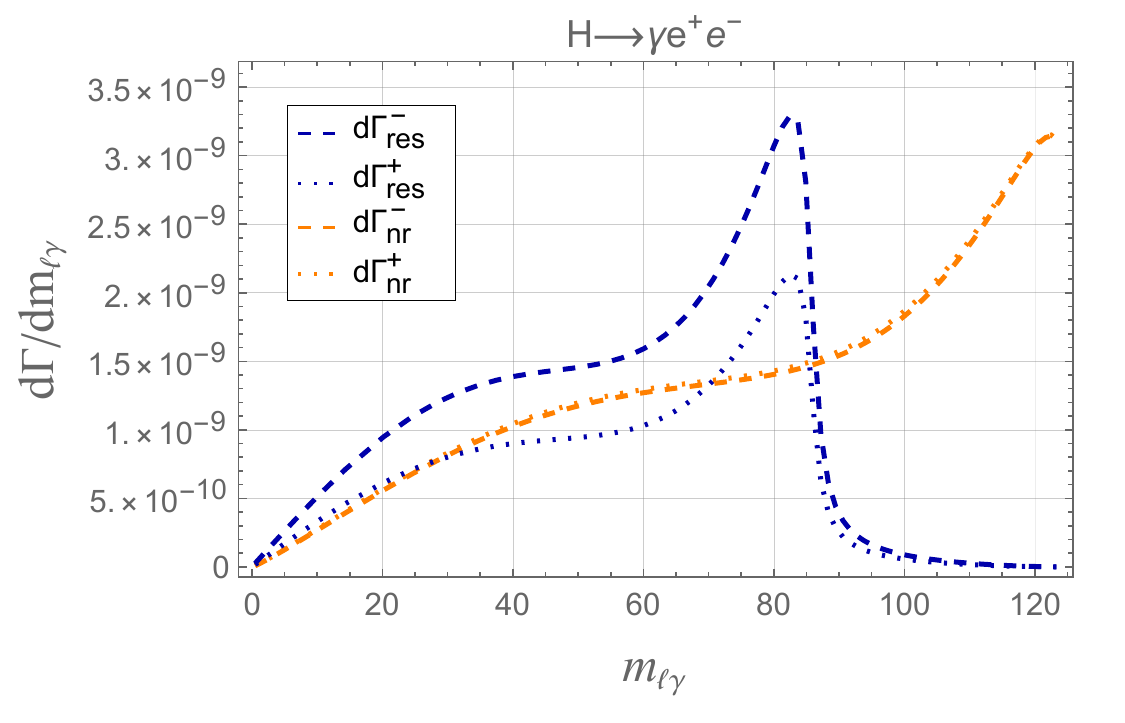}&
\includegraphics[width=3in,height=2in]{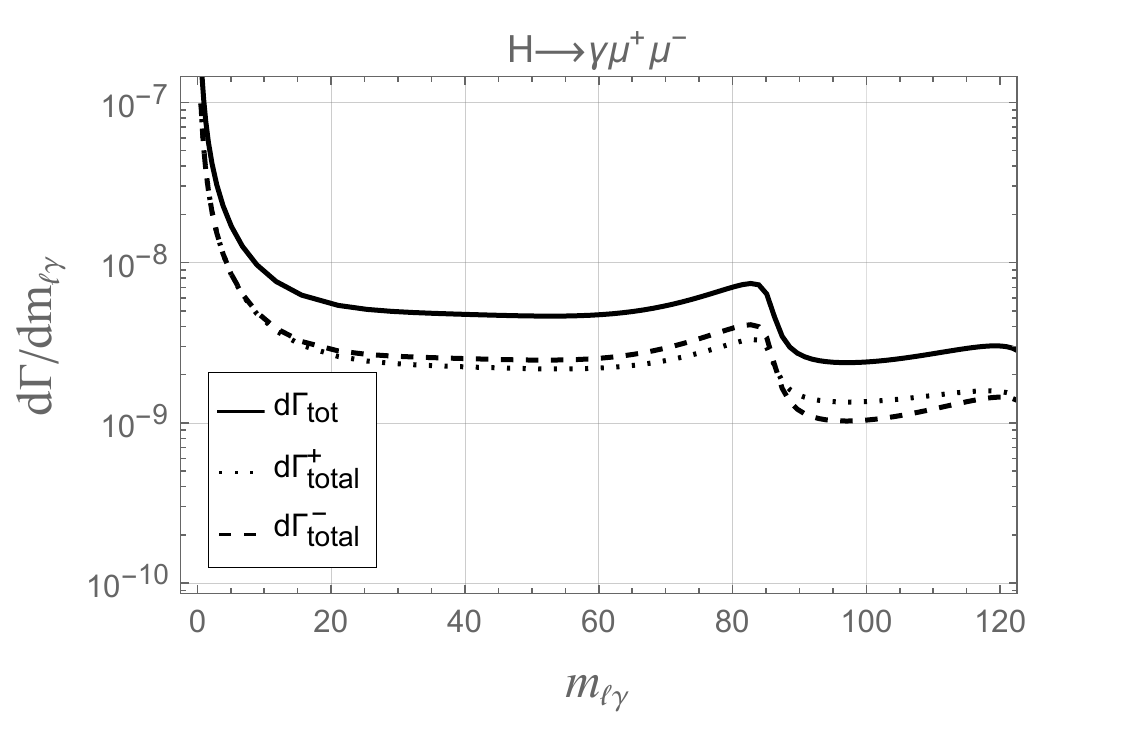}&\\
(b)&(e) \\
\includegraphics[width=3in,height=2in]{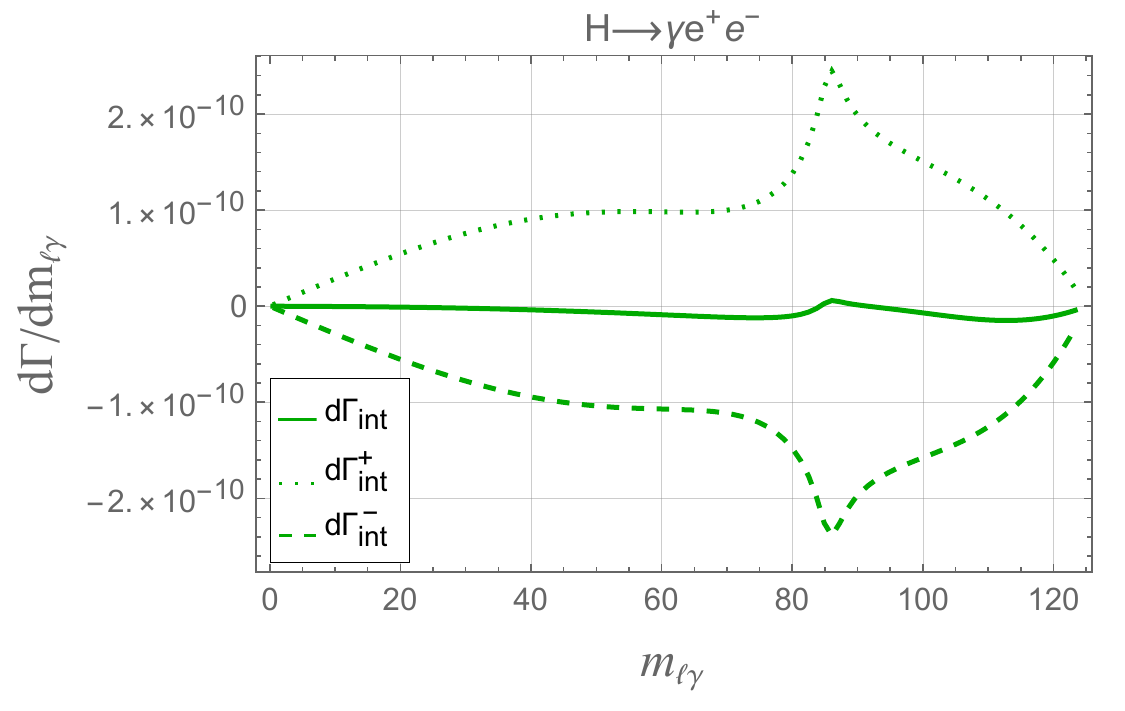}&
\includegraphics[width=3in,height=2in]{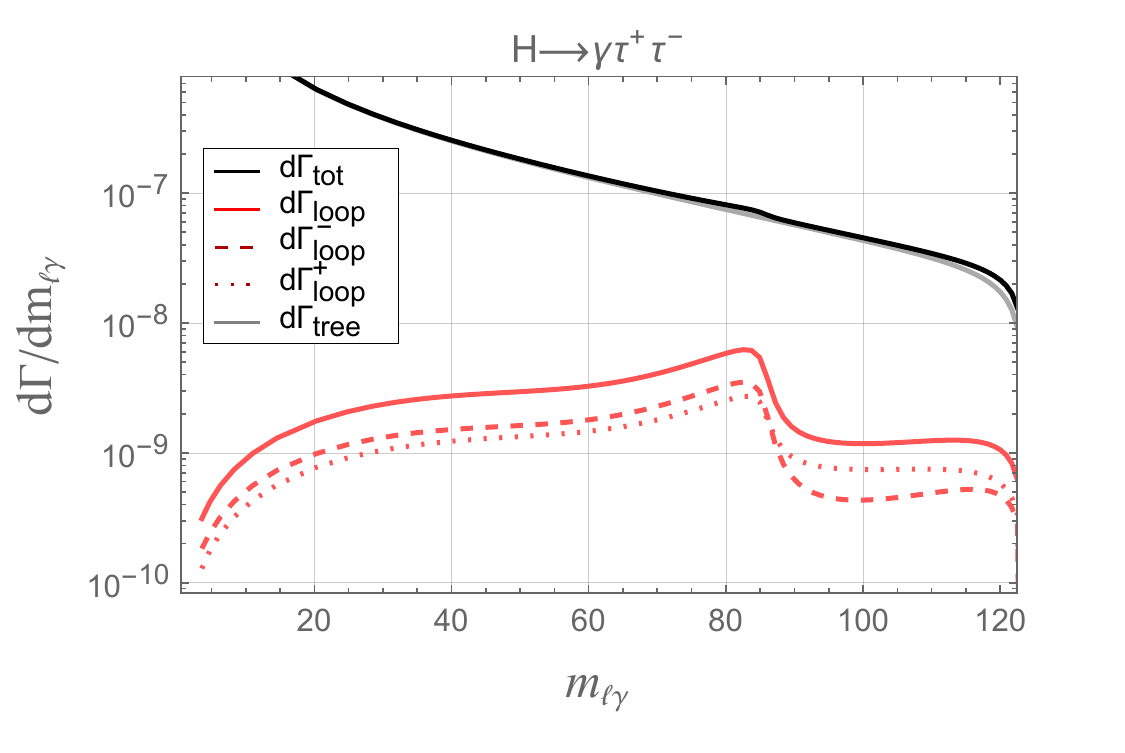}&\\
(c) & (f)
\end{tabular}
\caption{Longitudinally polarized differential decay rate with respect to invariant dilepton mass $m_{\ell\gamma}$. (a,d,f) show  unpolarized total, loop and tree contributions for $\ell=e,\mu,\tau$ respectively. Polarized parts of loop are also shown. (b) shows polarized resonance and non-resonance terms while (c) shows the polarized and unpolarized interference term for $\ell=e$. (e) shows the polarized as well as unpolarized total decay rate for $\ell=\mu$. The color scheme is same as described in Fig. \ref{LES}}
\label{LET}
\end{figure}

In Fig. \ref{LES}(d-e), we present differential decay rates $d\Gamma(m_{{\mu}{\mu}})$ for the  case of muon as a final state lepton. In contrast to the case of electron, the tree-level contribution is comparable to the loop contribution, $d\Gamma_{loop}\sim d\Gamma_{tree}$, especially, in the kinematic region between photon and Z pole. Hence, the total decay rate receives contributions from both tree and loop diagrams except at low $m_{\mu\mu}$ region (near the photon pole), where the tree level contribution is negligible. However, beyond the Z pole, the tree level contribution is dominant over the loop level contribution till the end of the spectrum.

 Around $m_{{\mu}{\mu}}=$ 60 Gev, we again see a sharp suppression of loop contribution to  $d\Gamma_{loop}^{-}$. However, the tree contribution $d\Gamma_{tree}^{-}$ raises the magnitude of the total polarized decay rate $d\Gamma_{tot}^{-}$ as shown in Fig. \ref{LES}(d). Therefore, the sharp dip (for electron) in $d\Gamma_{loop}^{-}$, gets smooth for the muon case as shown in the Fig. \ref{LES}(e). However, there is still around an order of magnitude difference in the total polarized decay rate $d\Gamma_{tot}^{\pm}$ for the final state muon between spin $-\frac{1}{2}$ and spin $+\frac{1}{2}$.

Due to the suppression of loop contribution to longitudinally polarized differential decay rate around $60$ GeV (see Fig. \ref{LES}(e), the role of muon Yukawa coupling with Higgs, $y_{\mu}=m_{\mu}/v$, becomes dominant in  differential decay rate when plotted against $m_{\mu\mu}$. This fact may provide an additional tool to measure Yukawa coupling of muon at lower energies. In fact, the Yukawa coupling is even more dominant near the Higgs pole, however that region is harder to explore experimentally. Furthermore, the 60 GeV region is particularly sensitive for probing new physics, since it is away from the photon, $Z$ and the Higgs poles.

The differential decay rates $d\Gamma(m_{{\tau}{\tau}})$ for the case of final state tauon are presented in Fig. \ref{LES}(f). Due to the large Higgs Yukawa coupling to tauons, $y_{\tau}=m_{\tau}/v$, the tree-level contribution dominates the total decay rate. However, the decay rate shows slightly different behavior near the photon pole. In this region, the tree level contribution of the total decay rate becomes comparable to the loop contribution. The same sharp decrease around $m_{\tau\tau}=60$ GeV is observed in the loop contribution of polarized decay rate $d\Gamma_{loop}^{-}$. However, for the tauon, the impact of this suppression is overshadowed by the large tree contribution.

Fig. \ref{LET}(a) describes the longitudinally polarized differential decay rate $d\Gamma(m_{e\gamma})$ against invariant electron-photon mass $m_{e\gamma}$. It is noted that before the Z pole, the longitudinally polarized decay rate $d\Gamma_{tot}^{-}$ is larger than $d\Gamma_{tot}^{+}$ due to the large resonance contribution $d\Gamma_{res}^{-}$ as shown in Fig. \ref{LET}(b,c). Beyond the Z pole,  the longitudinally polarized decay rate $d\Gamma_{tot}^{+}$ becomes larger than $d\Gamma_{tot}^{-}$. In this region, the non-resonance contribution $d\Gamma_{nr}^{\pm}$ becomes dominant over resonance contribution $d\Gamma_{res}^{\pm}$ and the interference term $d\Gamma_{res-nr}^{\pm}$. However, there is no considerable difference with respect to spin polarization i.e., $d\Gamma_{res(nr)}^{+}\simeq d\Gamma_{res(nr)}^{-}$ as shown in Fig. \ref{LET}(b). But the interference term $d\Gamma_{res-nr}^{\pm}$ is polarization dependent with opposite signs $d\Gamma_{res-nr}^{-}\simeq -d\Gamma_{res-nr}^{+}$ as can be seen in Fig. \ref{LET}(c). The negative contribution $d\Gamma_{res-nr}^{-}$ makes $d\Gamma_{tot}^{-}$ smaller than the $d\Gamma_{tot}^{+}$ beyond the Z pole, as shown in Fig. \ref{LET}(a).

Fig. \ref{LET}(d) shows the longitudinally polarized decay rate for the final state muon against $m_{\mu\gamma}$. Near the lower $m_{\mu\gamma}$ region, we observe that the tree contribution dominates. In the rest of the region, the loop contribution is dominant, where the behavior is similar to that of the electron. However, for tauon, the tree level contribution dominates throughout the whole kinematic region, therefore the difference in differential decay rate for polarized tauon is negligible as shown in Fig. \ref{LET}(f).

\subsubsection{Normally polarized differential decay rates}

\begin{figure}
\centering
\begin{tabular}{cc}
\includegraphics[width=3in,height=2in]{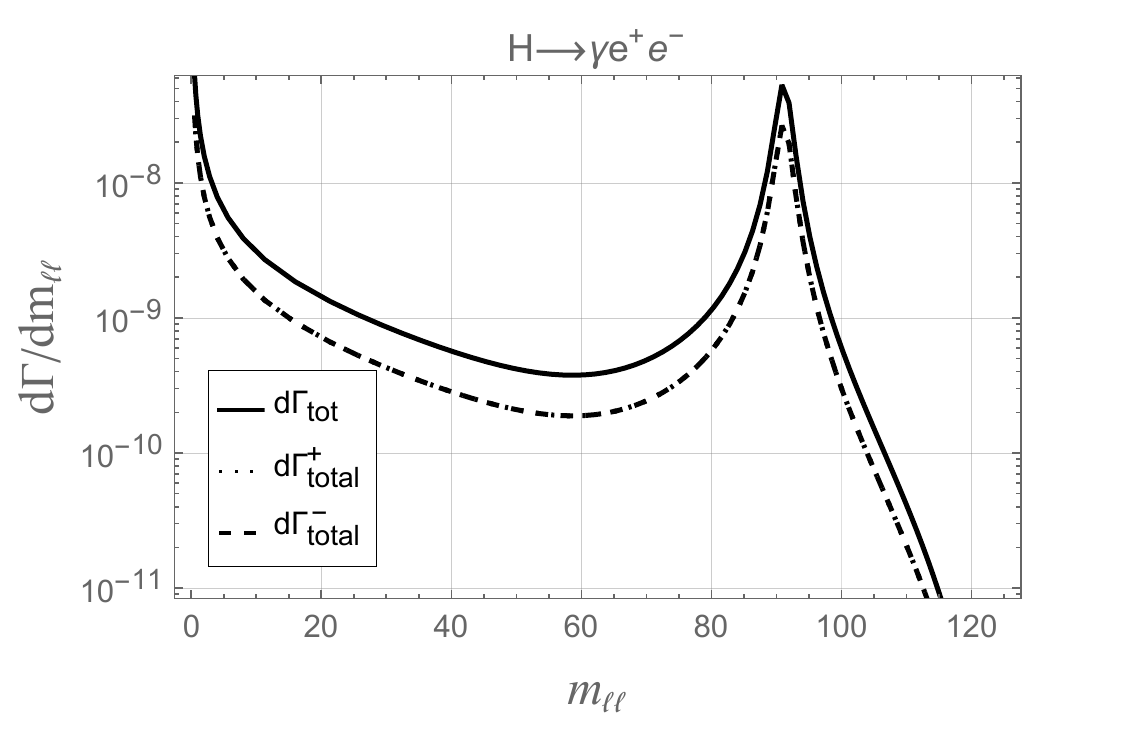}&
\includegraphics[width=3in,height=2in]{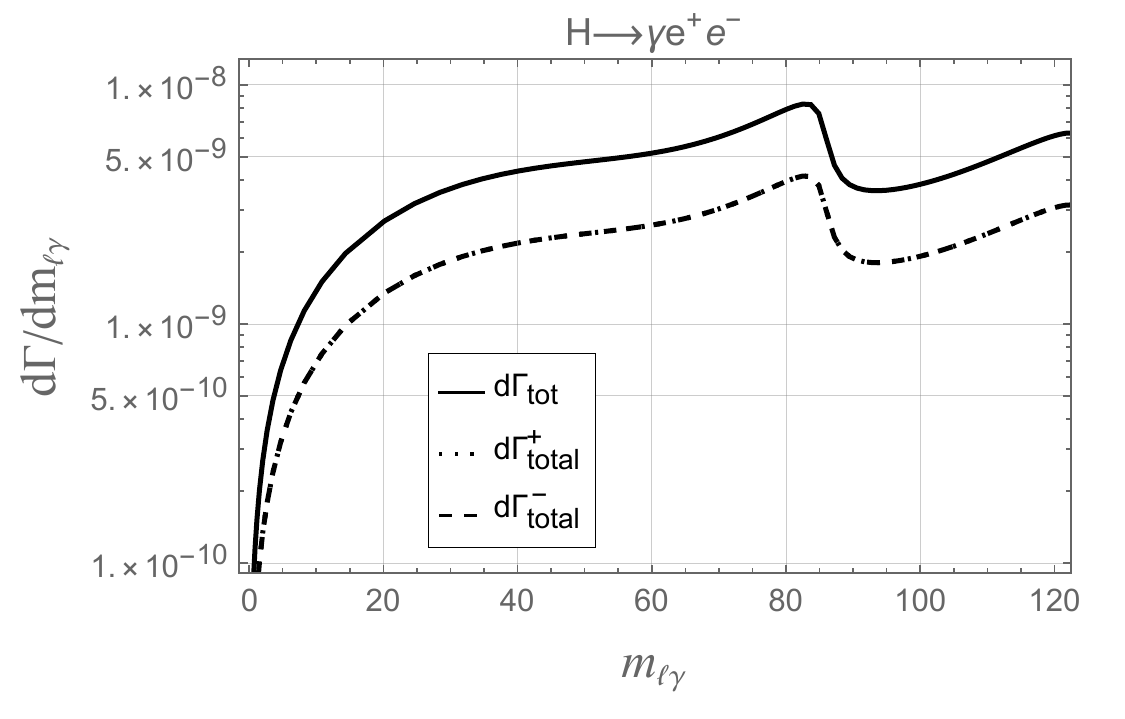}\\
(a) & (d) \\
\includegraphics[width=3in,height=2in]{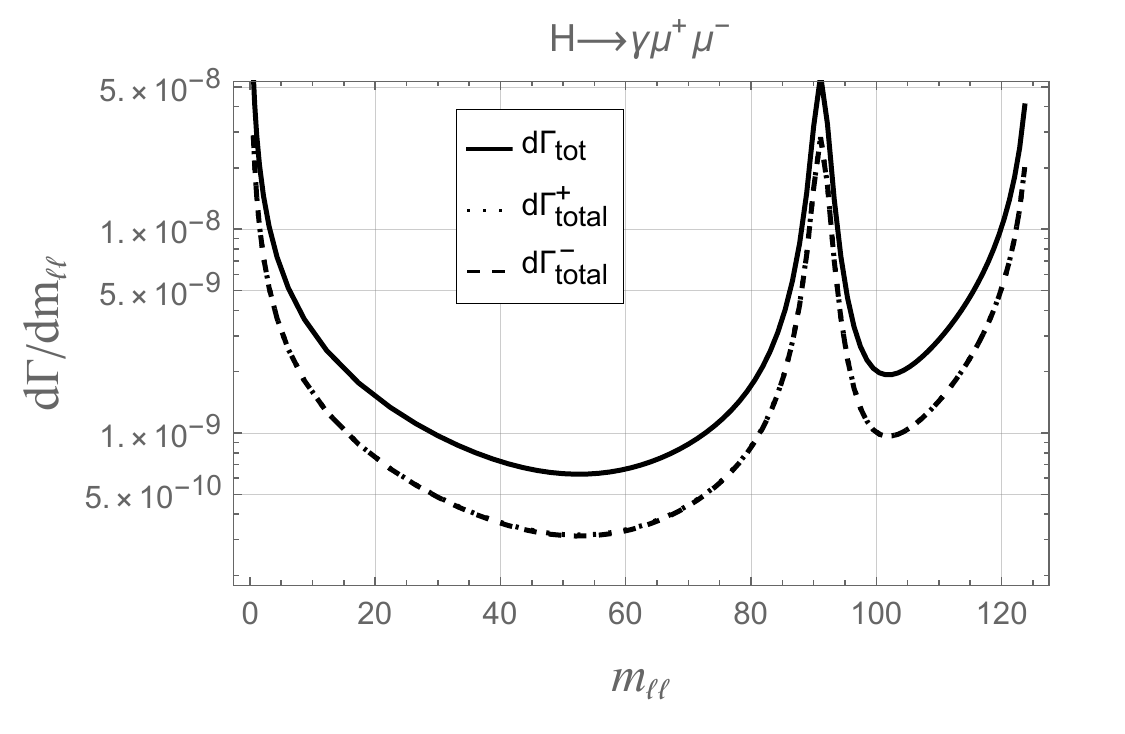}&
\includegraphics[width=3in,height=2in]{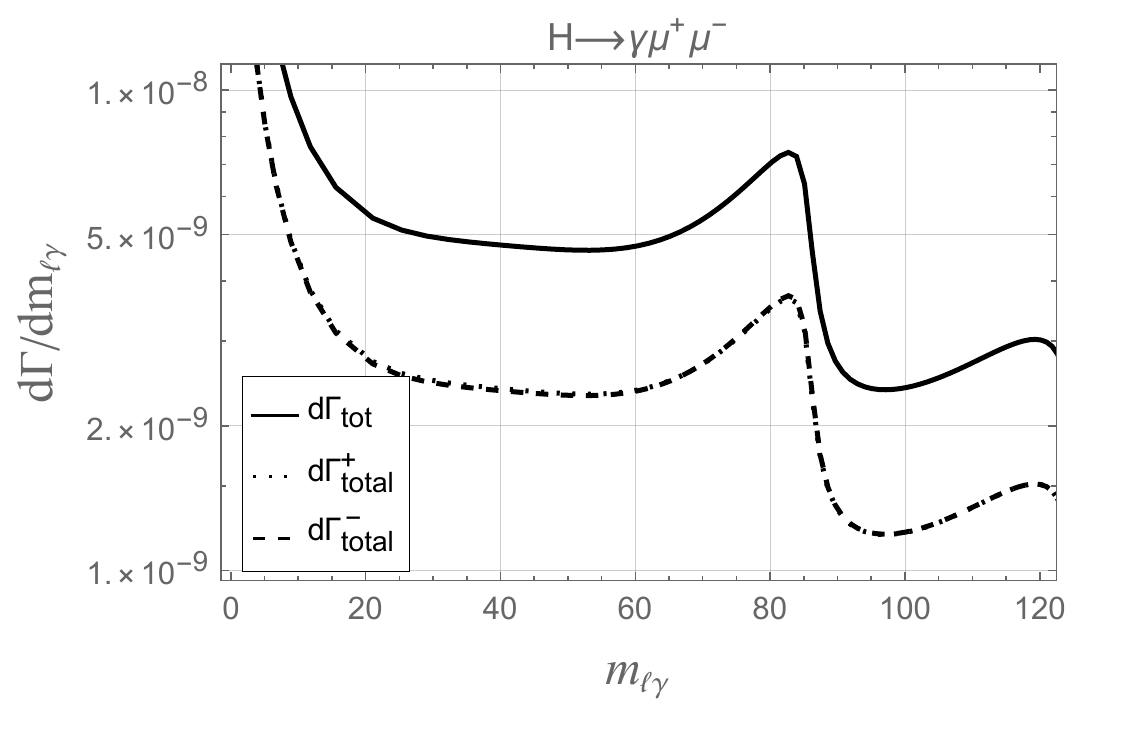}\\
(b) & (e) \\
\includegraphics[width=3in,height=2in]{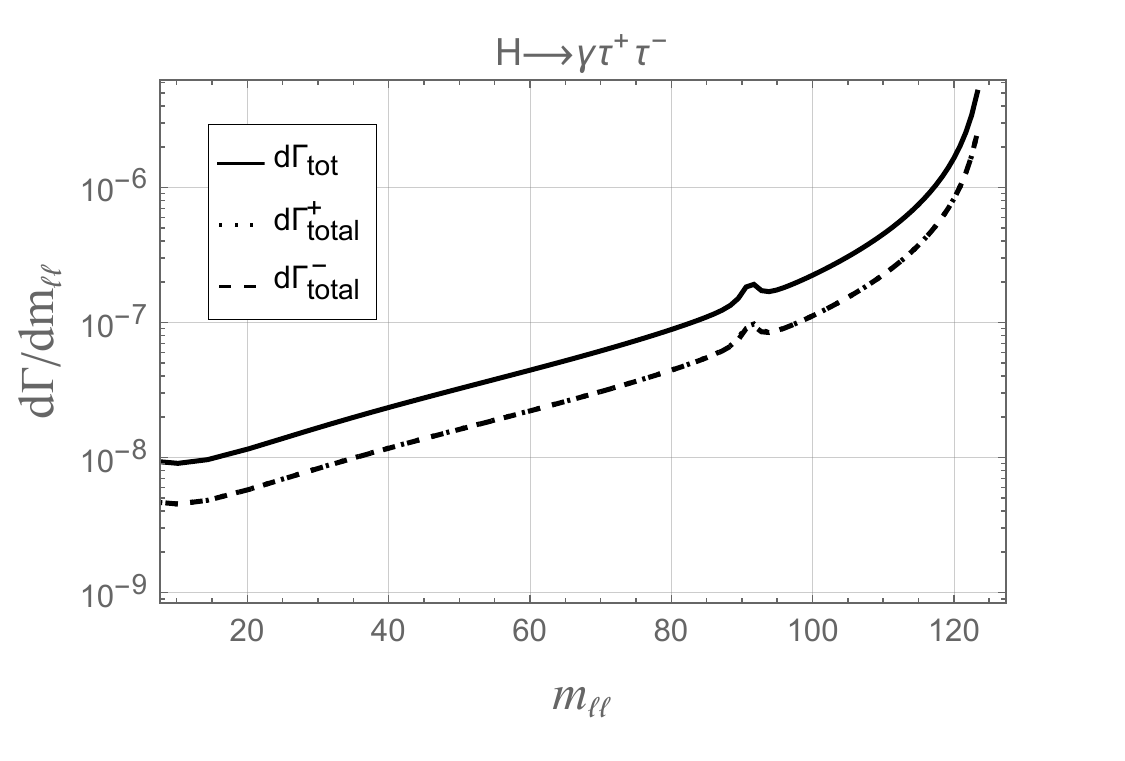}&
\includegraphics[width=3in,height=2in]{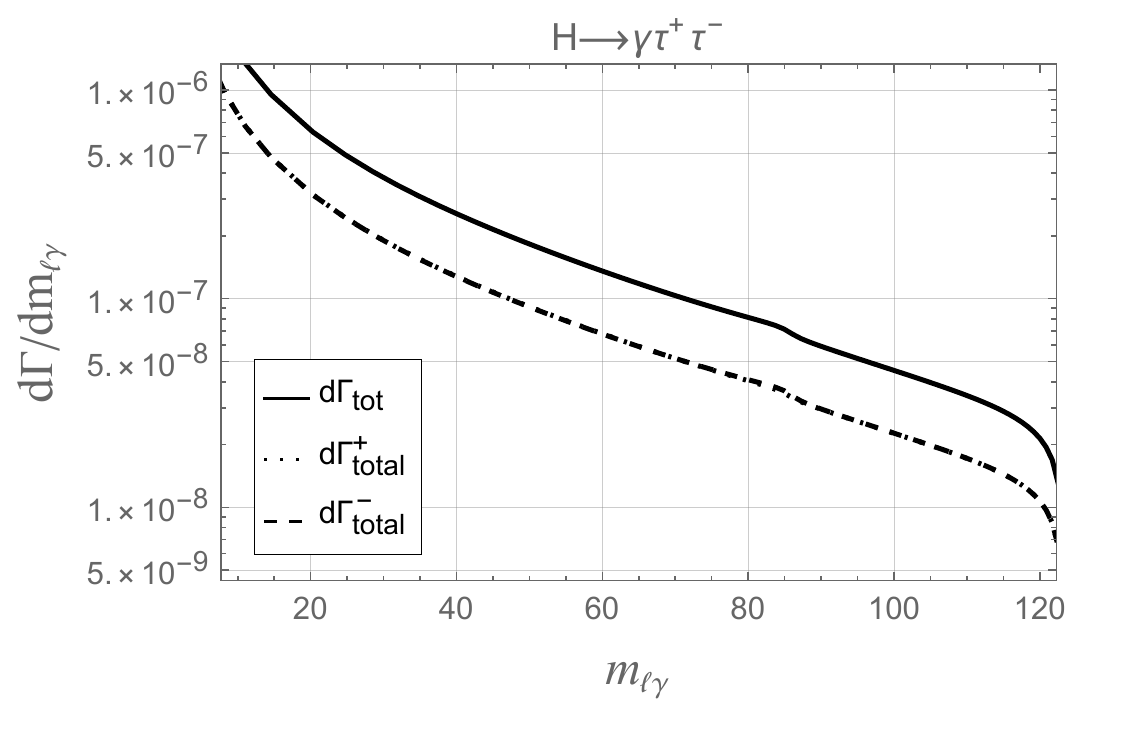}\\
(c) & (f)
\end{tabular}
\caption{Normally polarized differential decay rate with respect to invariant masses $m_{\ell\ell}$ and $m_{\ell\gamma}$ for all leptons. First(second) column shows $m_{\ell\ell}$ ($m_{\ell\gamma}$) plots  while the three rows show respective leptons $\ell=e,\mu,\tau$.}
\label{normal}
\end{figure}

The plots of total decay rates against $m_{\ell\ell}$ and $m_{\ell\gamma}$ for normally polarized electron, muon and tauon are shown in Fig. \ref{normal} in respective rows. In each case, both spin polarizations of final state leptons almost equally contribute to the corresponding total differential decay rates $d\Gamma_{tot}^{+}\simeq d\Gamma_{tot}^{-}$.

\begin{figure}
\centering
\begin{tabular}{cc}
\includegraphics[width=3in,height=2in]{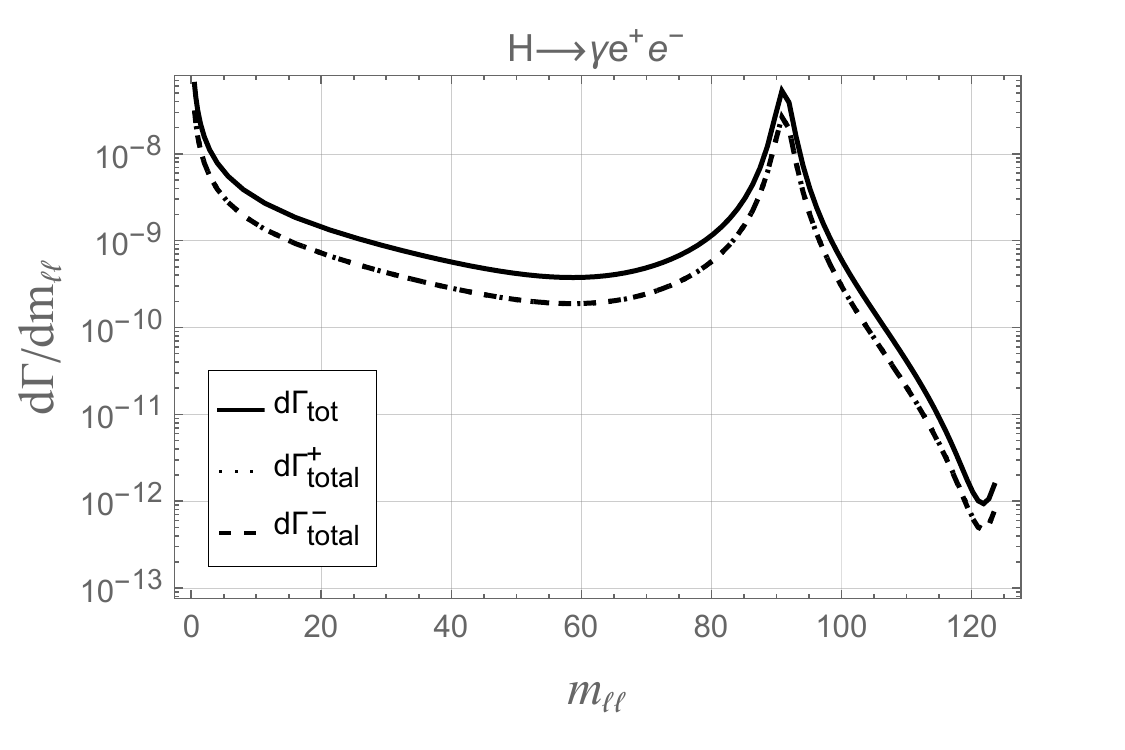}&
\includegraphics[width=3in,height=2in]{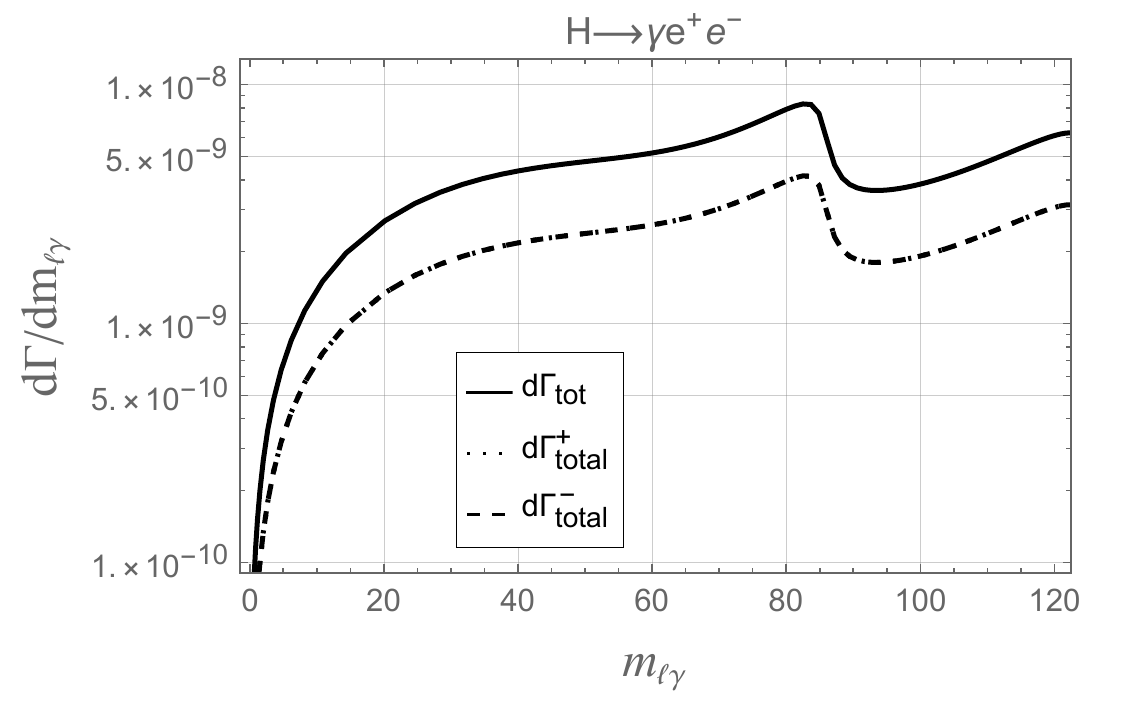}\\
(a) & (d) \\
\includegraphics[width=3in,height=2in]{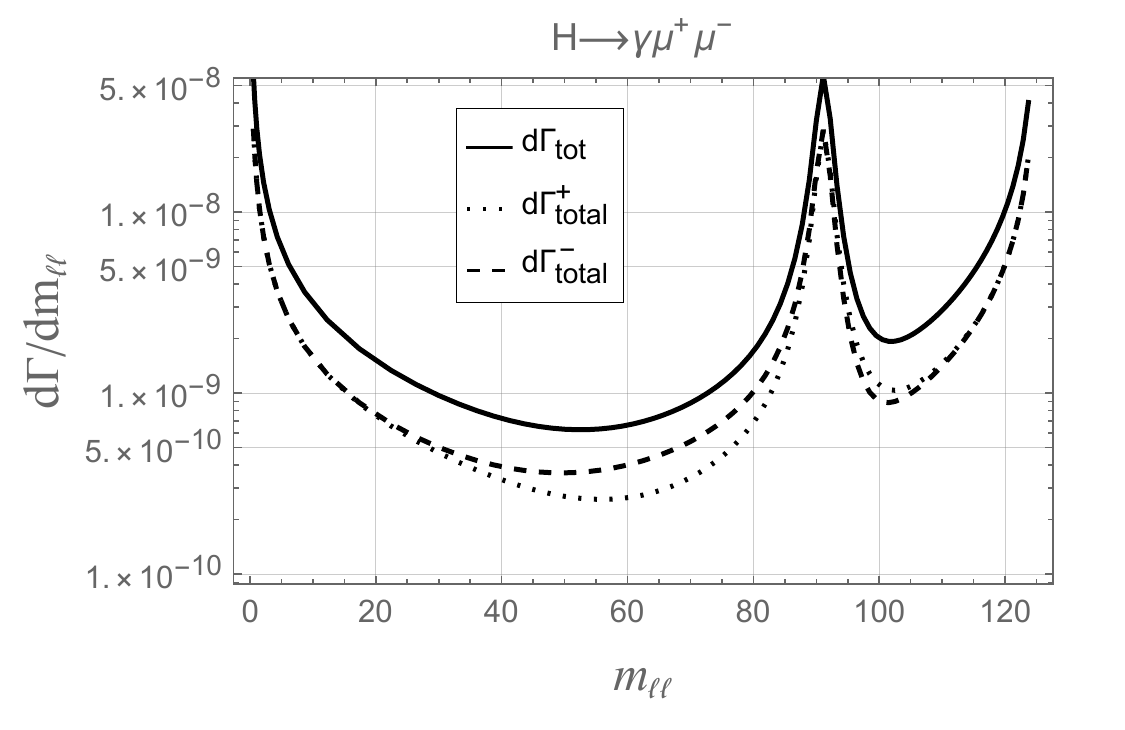}&
\includegraphics[width=3in,height=2in]{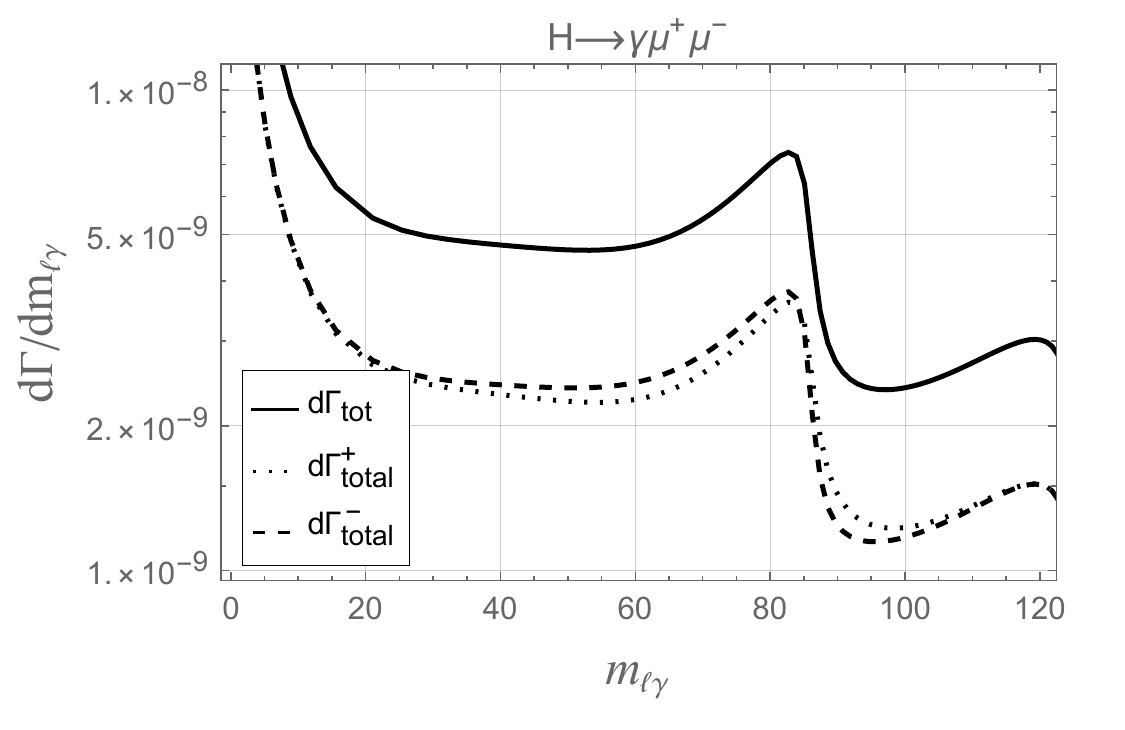}\\
(b) & (e) \\
\includegraphics[width=3in,height=2in]{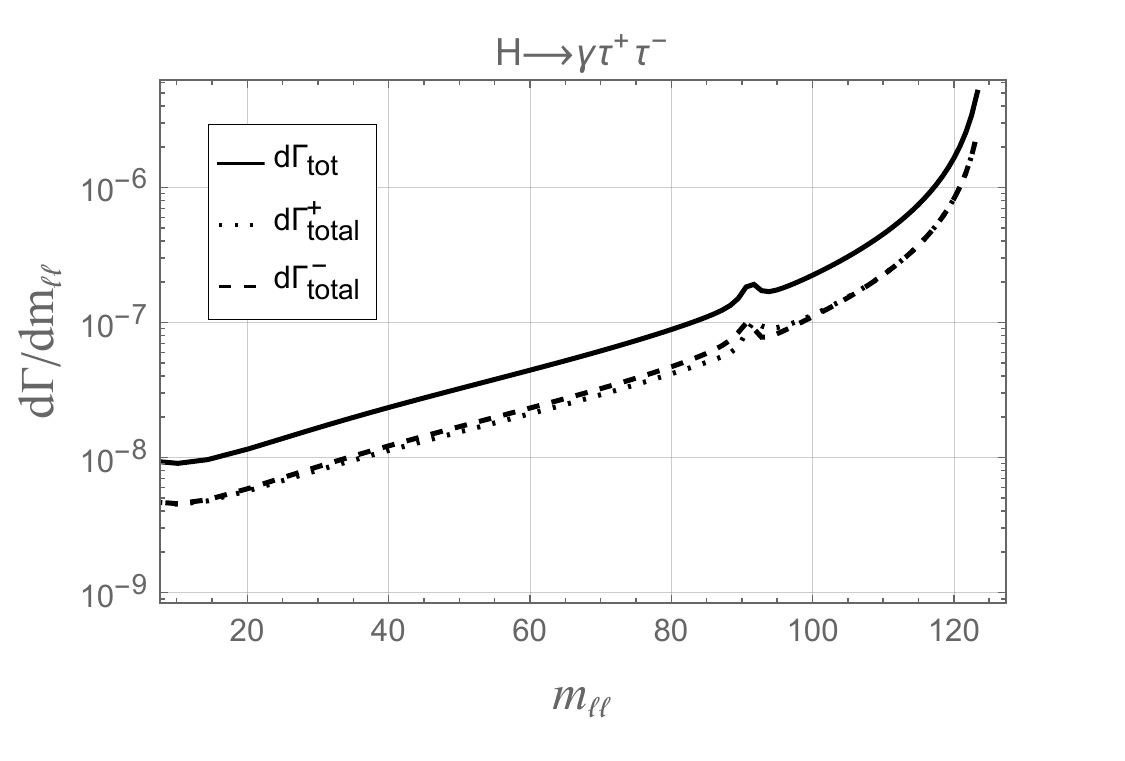}&
\includegraphics[width=3in,height=2in]{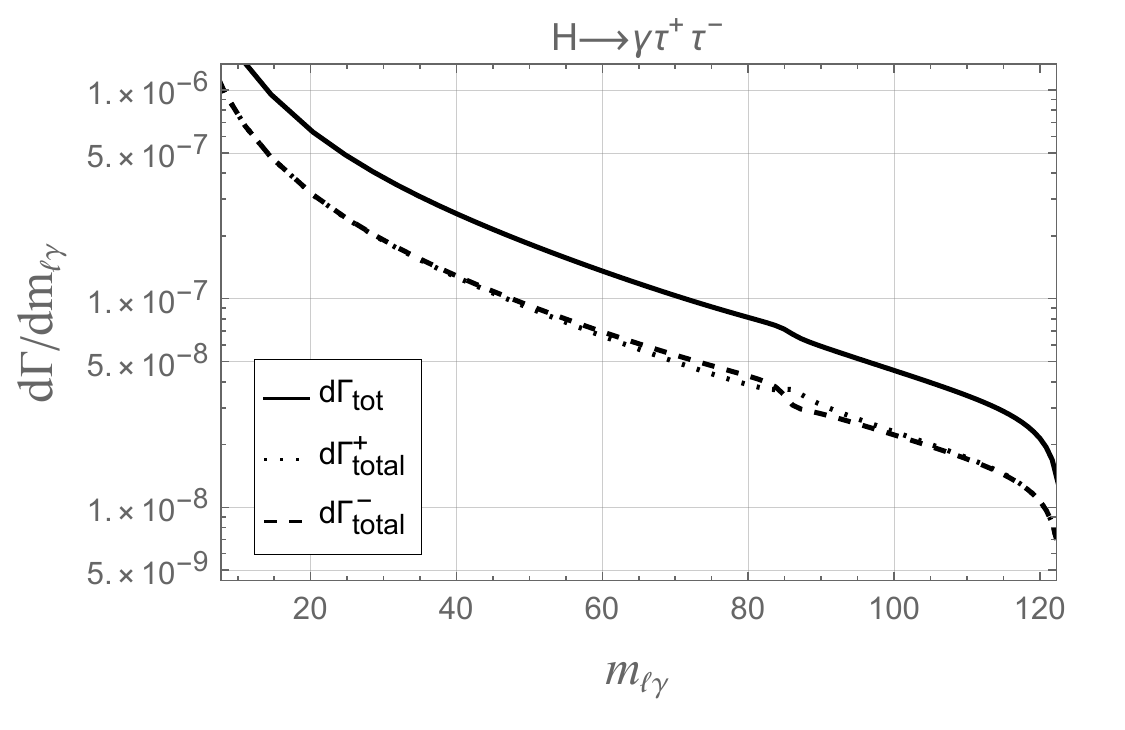}\\
(c) & (f)
\end{tabular}
\caption{Transverse polarized differential decay rate with respect to invariant masses $m_{\ell\ell}$ and $m_{\ell\gamma}$ for all leptons. First(second) column shows $m_{\ell\ell}$ ($m_{\ell\gamma}$) plots  while the three rows show respective leptons $\ell=e,\mu,\tau$.}
\label{transverse}
\end{figure}

\begin{figure}
\centering
\begin{tabular}{cc}
\includegraphics[width=3in,height=2in]{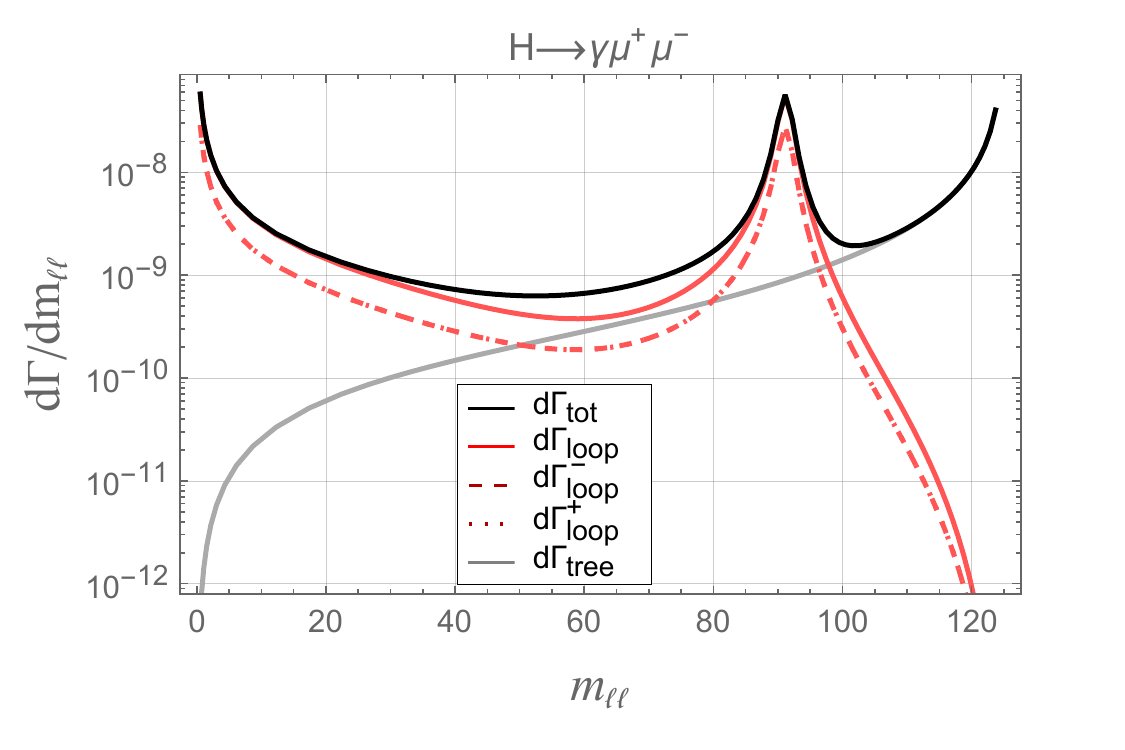}&
\includegraphics[width=3in,height=2in]{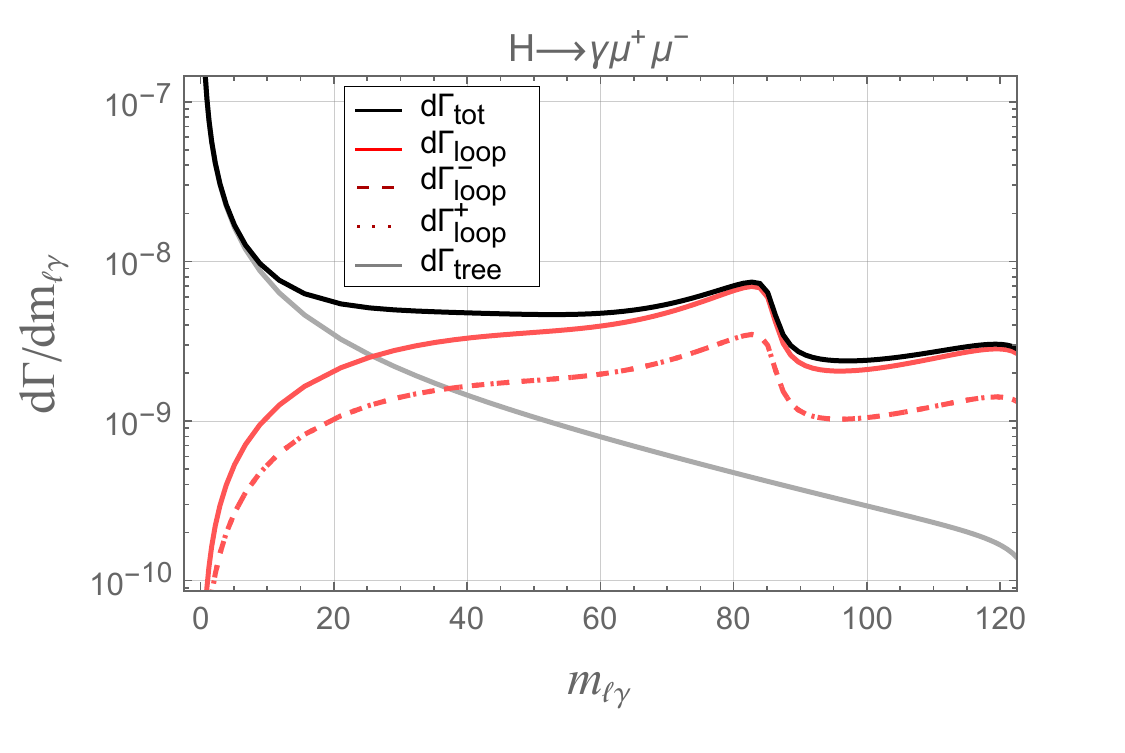}\\
(a) & (d) \\
\includegraphics[width=3in,height=2in]{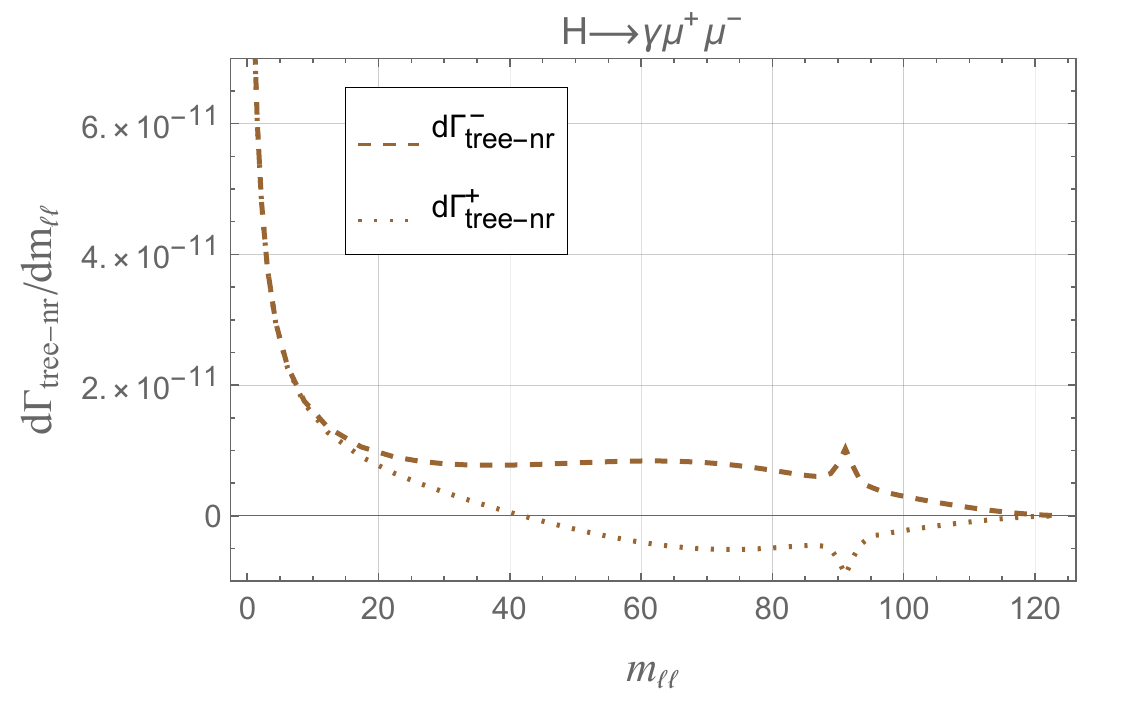}&
\includegraphics[width=3in,height=2in]{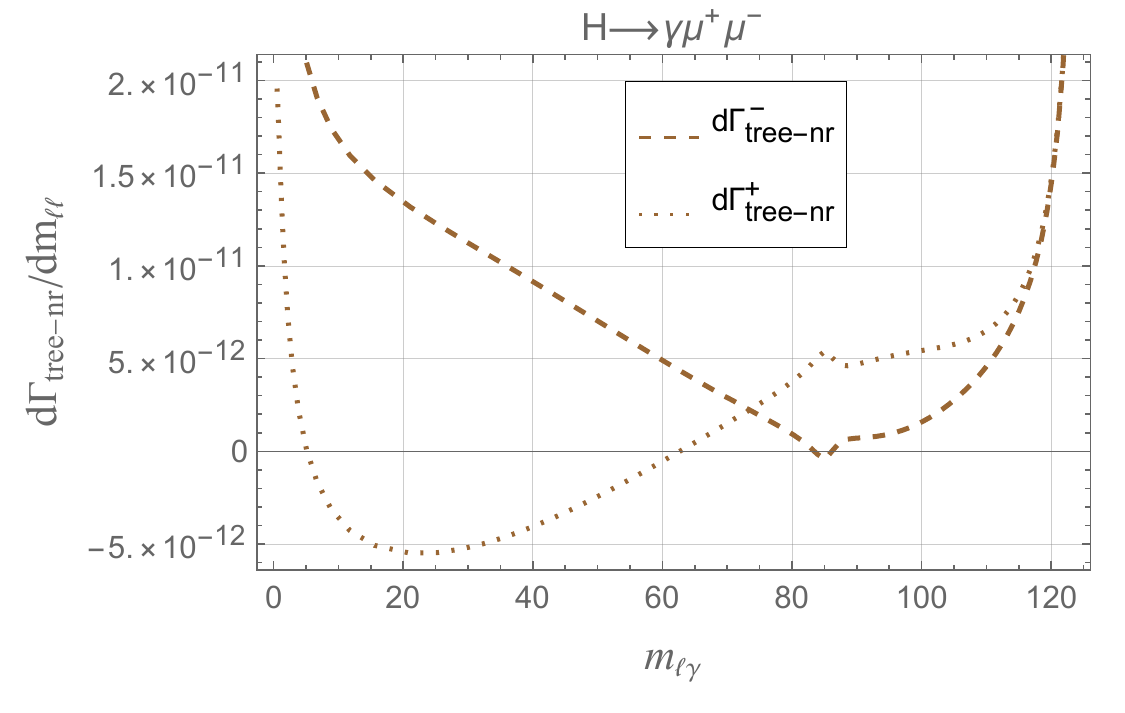}\\
(b) & (e) \\
\includegraphics[width=3in,height=2in]{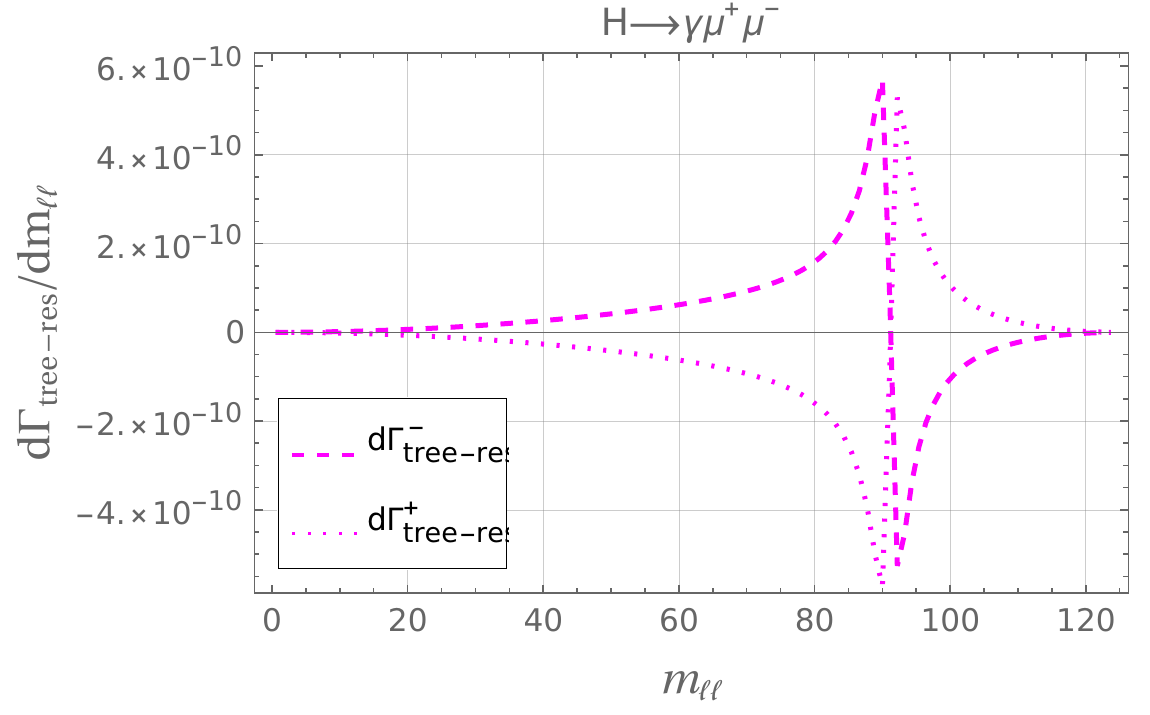}&
\includegraphics[width=3in,height=2in]{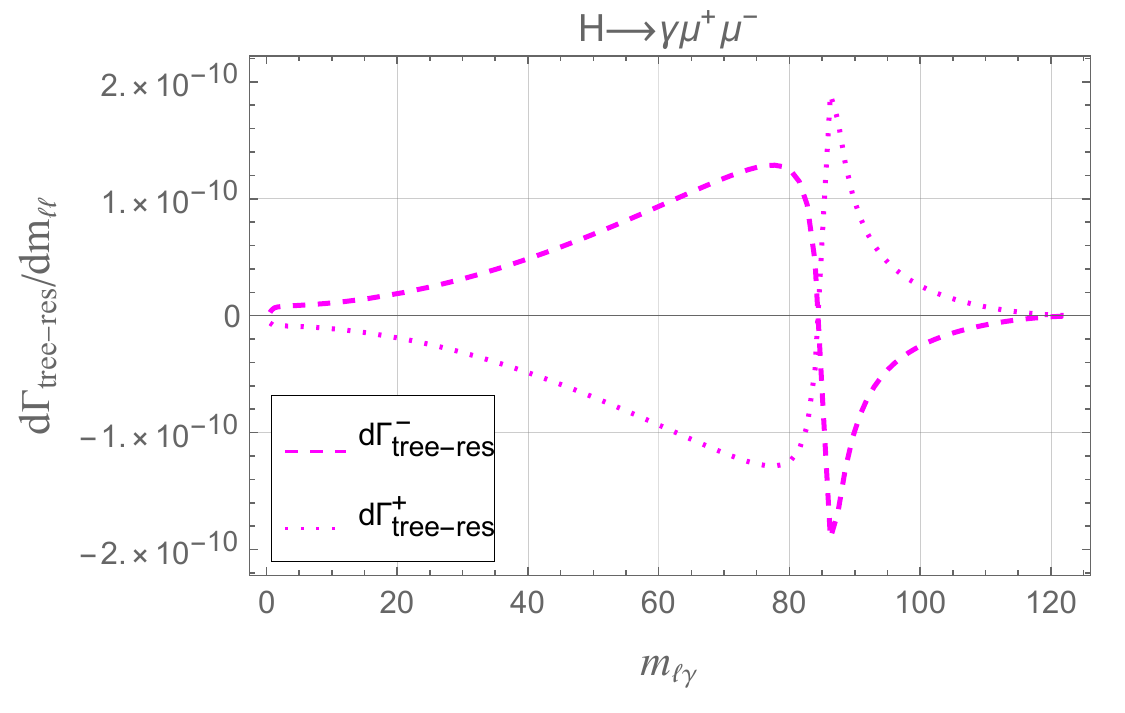}\\
(c) & (f)
\end{tabular}
\caption{Transverse polarized differential decay rate with respect to invariant masses $m_{\mu\mu}$ and $m_{\mu\gamma}$ is shown in respective columns. First row shows unpolarized rates at all levels and polarized at loop level. Second row shows tree and non-resonance interference term of polarized rate while third row shows tree and resonance interference term of polarized rate.\textbf{ Color scheme:} The solid lines show $d\Gamma$ while $d\Gamma^{\pm}$ for the final state lepton with polarization $+\frac{1}{2}$ and $-\frac{1}{2}$ are represented by dotted and dashed lines respectively. $d\Gamma_{tot}$, $d\Gamma_{tree}$ and $d\Gamma_{loop}$ are  represented by the black, gray and red lines respectively. The brown line shows $d\Gamma_{tree-nr}$ while the magenta represents $d\Gamma_{tree-res}$.}
\label{TMTL}
\end{figure}

\subsubsection{Transversely polarized differential decay rates}

Fig. \ref{transverse} contains the total differential decay rates for transversely polarized final state leptons. We observe that the case of transversely polarized electron and tauon is similar to that of normally polarized ones and no notable difference is found for both spin polarizations. 

However, transversely polarized muon decay rate behaves differently than the normally polarized case. Before the Z pole, around 60 GeV, the decay rate of transversely polarized muon with spin $-\frac{1}{2}$ is larger than that of muon with spin $+\frac{1}{2}$, while after the Z pole, the opposite behavior is observed as shown in the plots against $m_{\mu\mu}$ and $m_{\mu\gamma}$ (Figs. \ref{transverse}(b) and \ref{transverse}(e)). Since, there is no such behavior in the case of electrons, we argue that the difference in the decay rate of spin polarized muon does not  receive contribution  from the loop. This fact leads us to analyze the tree and loop interference term to investigate the decay rate for muon case. Therefore, we have shown the different important contributions to decay rates for transversely polarized final state muon in the Fig. \ref{TMTL}. One can notice that polarized decay rates at the loop level, $d\Gamma^{\pm}_{loop}$ behave in the same way for different polarizations as shown in first row of Fig. \ref{TMTL}. The last two rows of Fig. \ref{TMTL} show interference of tree contribution with resonance and non-resonance contributions, $d\Gamma^{\pm}_{tree-res}$ and $d\Gamma^{\pm}_{tree-nr}$, respectively. The contribution of these terms to the total polarized decay rate is similar for $m_{\ell\ell}$ and $m_{\ell\gamma}$ plots. Hence, the following arguments apply equally to both cases. Fig. \ref{TMTL} shows that these interference terms behave differently for different spin polarizations, in contrast to the behavior of  $d\Gamma^{\pm}_{loop}$.

We further observe that the term $d\Gamma^{\pm}_{tree-nr}$ is much smaller in magnitude as compared to $d\Gamma^{\pm}_{tree-res}$ term in the region before the Z pole, around 60 GeV, as well as after the Z pole. This is the region of particular interest as the difference in decay rate for $d\Gamma_{tot}^{+}$ and $d\Gamma_{tot}^{-}$ is significant as shown in Fig. \ref{transverse}(b,e). Additionally, the interference term,  $d\Gamma_{tree-res}^+$  is negative whereas, $d\Gamma_{tree-res}^-$ is positive in the region of interest as shown in Fig. \ref{TMTL}(c,f) with considerable magnitude. As a consequence, the total polarized decay rate $d\Gamma_{tot}^{+}$ is smaller as compared to $d\Gamma_{tot}^{-}$ before the Z pole. However, after the Z pole, the above contributions to polarized decay rate change signs for both the spin polarizations, hence the opposite behavior is observed, as shown in Fig. \ref{transverse}(b,e). This concludes our discussion about polarized and unpolarized decay rates.

\subsection{Lepton polarization asymmetry $P_{i}$}

\begin{figure}
\centering
\begin{tabular}{cc}
\includegraphics[width=3in,height=2in]{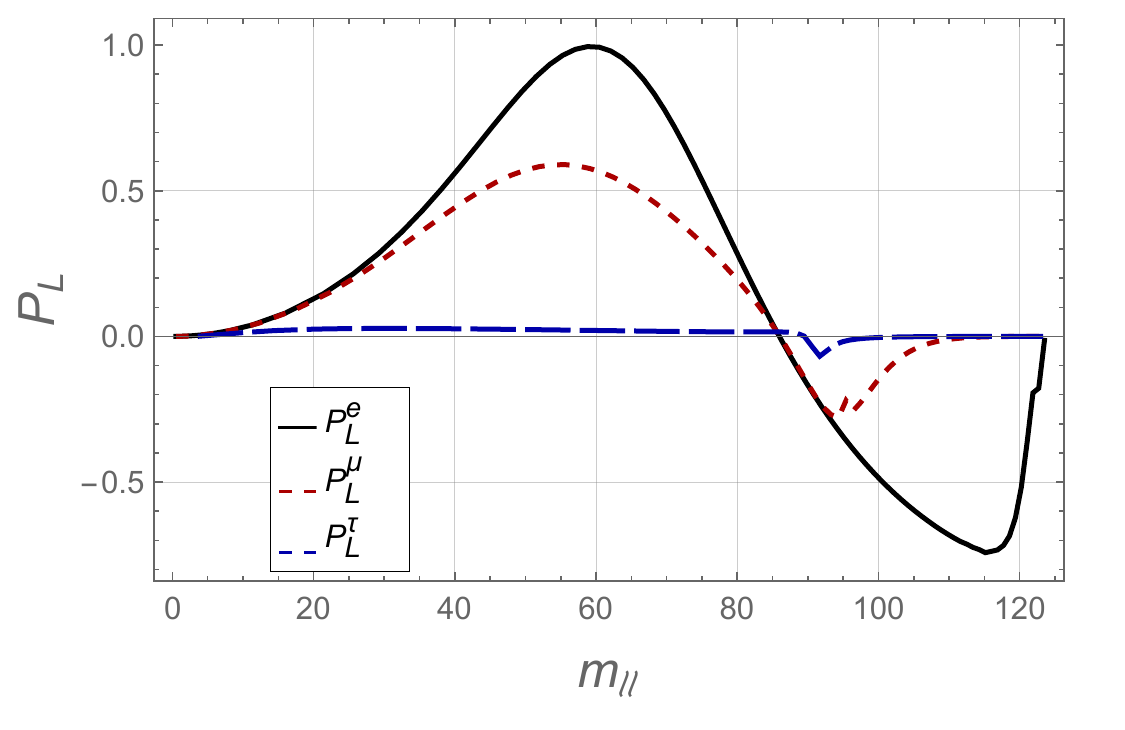}&
\includegraphics[width=3in,height=2in]{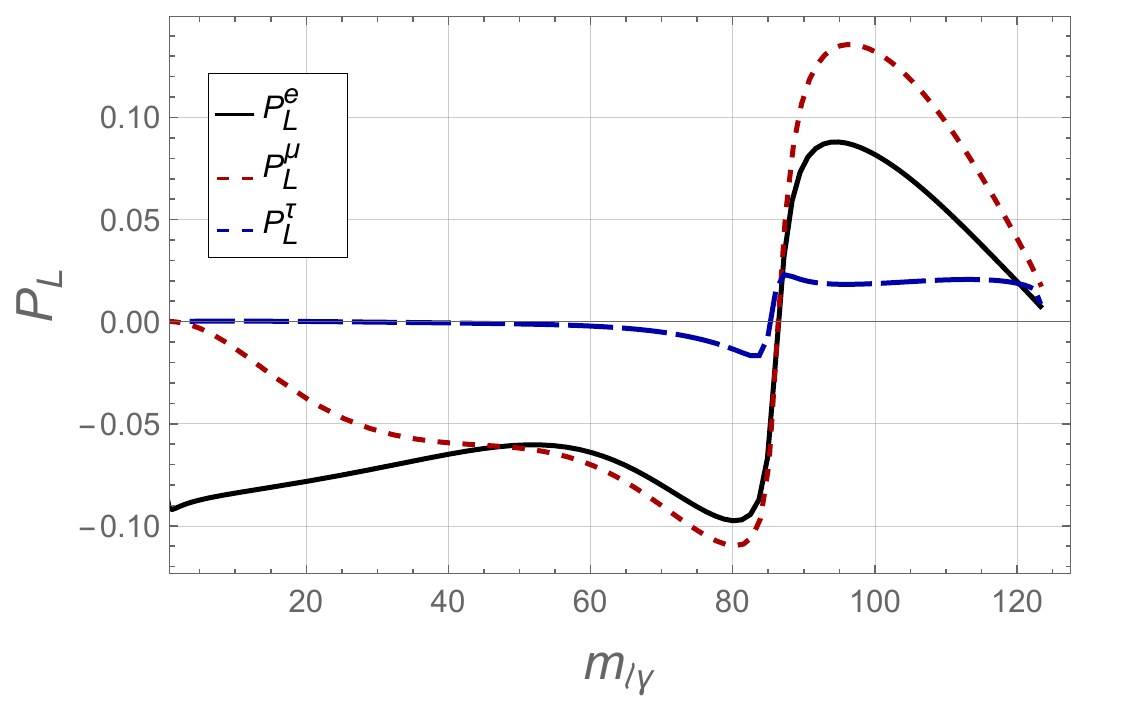}\\
(a)&(b)\\
\includegraphics[width=3in,height=2in]{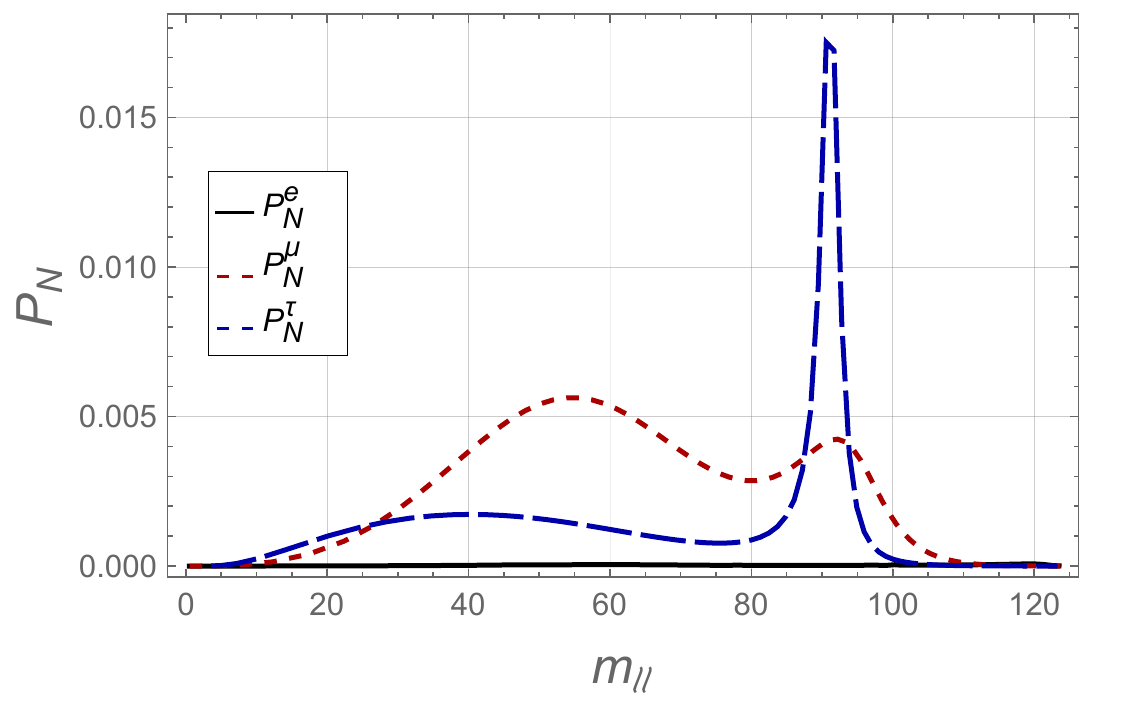}&
\includegraphics[width=3in,height=2in]{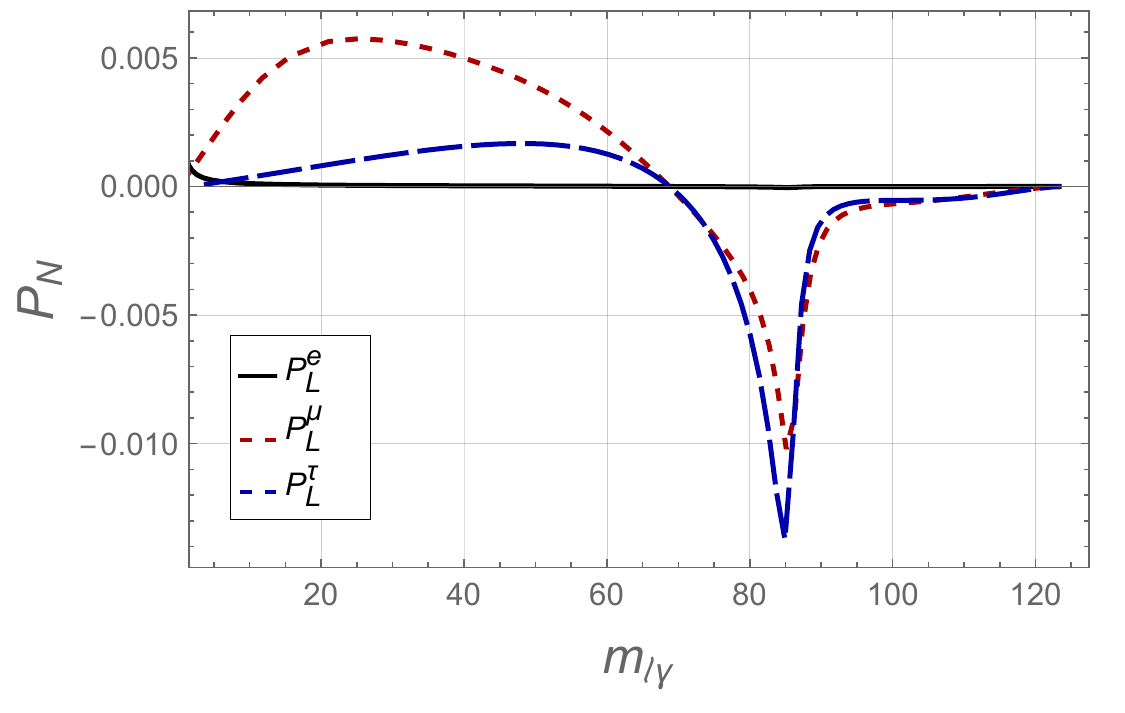}\\
(c)&(d)\\
\includegraphics[width=3in,height=2in]{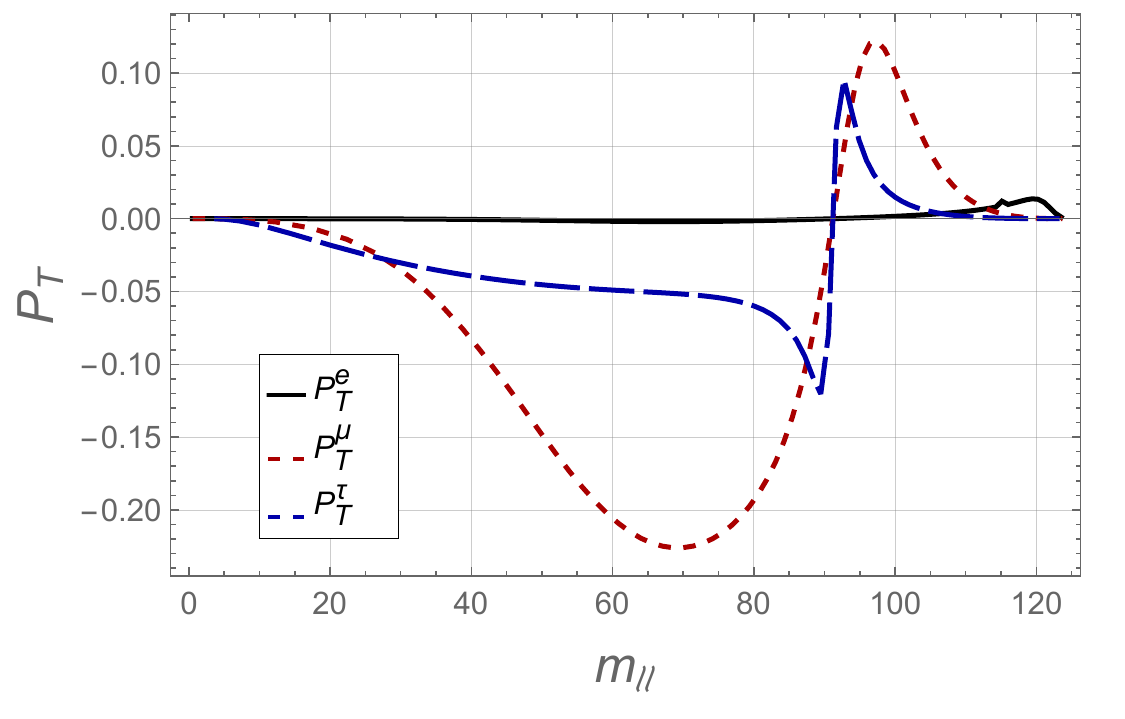}&
\includegraphics[width=3in,height=2in]{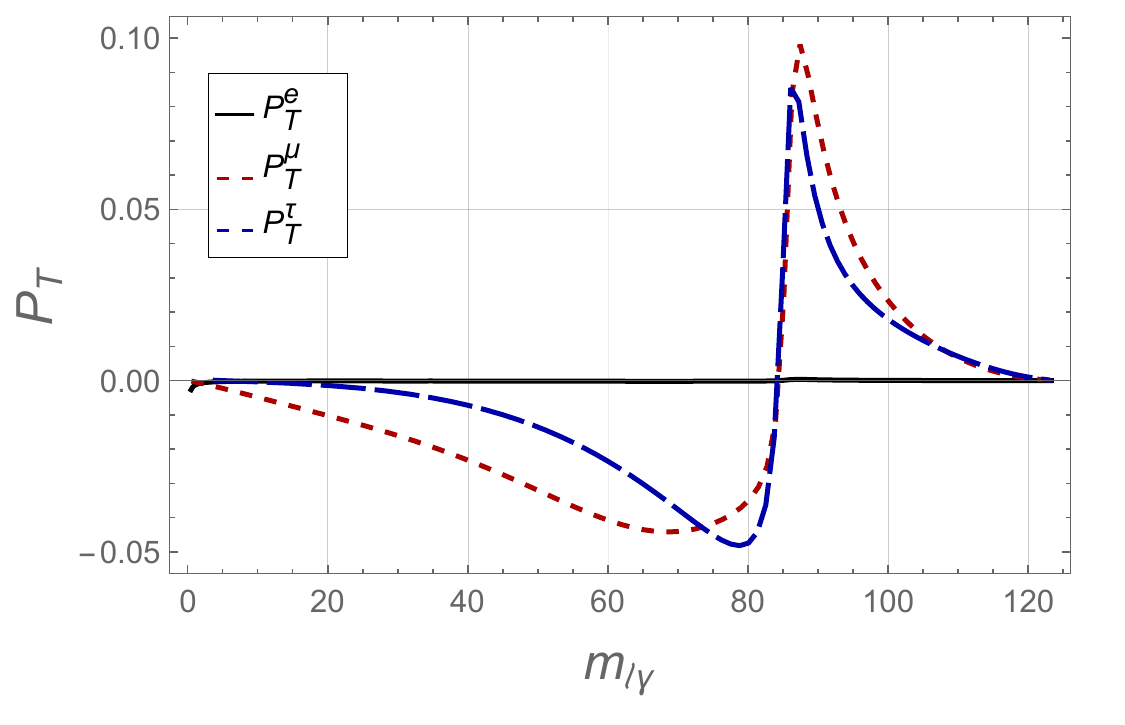}\\
(e)&(f)
\end{tabular}
\caption{Polarization asymmetry $P_i$ where $i=L,N,T$ represents longitudinal, normal and transverse polarization for all $\ell=e,\mu,\tau$. \textbf{Color Scheme:} Black solid line represents $P_i$ when electron is polarized, red dotted and blue dashed lines represent case of muon and tauon respectively.}
\label{PAS}
\end{figure}

In this section, we analyze other notable observables i.e., spin polarization asymmetries, $P_{L,T,N}$, for longitudinal, transverse and normal polarizations, respectively. These polarization asymmetries, defined in Eqn. (\ref{singlepol}), are plotted against $m_{\ell\ell}$ as well as $m_{\ell\gamma}$ in Fig. \ref{PAS}. Solid black curve represents final state electron, red dotted curve represents muon and blue dashed line represents tauon. Numerical values of longitudinal and transverse polarization asymmetries for different bins of $m_{\ell\ell}$ are given in Table \ref{wc table}, whereas, normal polarization asymmetries are less than 1$\%$ and therefore not given in the table.

All polarization asymmetries change sign at Z pole except $P_N(m_{\ell\ell})$, which remains positive throughout the whole kinematic region. We observe that the zero-crossings occur at Z pole except for $P_N(m_{\ell\gamma})$ for which the zero-crossing shifts slightly left from the Z pole. Moreover, $P_N(m_{\ell\gamma})$  and  $P_L(m_{\ell\ell})$ change signs from positive to negative, while rest of polarization asymmetries $P_L (m_{\ell\gamma})$ , $P_T(m_{\ell\ell})$ and $P_T(m_{\ell\gamma})$ change their signs from negative to positive. In the coming subsections, we discuss different generations of leptons separately.

\begin{table*}
\scalebox{0.9}{
\begin{tabular}{c|c|c|c|c|c|cc}
\hline\hline
$s_{min}-s_{max}(\text{GeV}^2)$&$10^2-30^2$&$30^2-50^2$&$50^2-70^2$&$70^2-90^2$&$90^2-110^2$&full
Phase space
\\ \hline
 \ \  $P_L(H\to\gamma e^+e^-)$ \ \  & \ \  11.3\% \ \  & \ \  53.3\% \ \  & \ \ 92.6\% \ \  & \ \  -0.4\% \ \  & \ \  -24.3\% \ \  & \ \  -4.8\% \ \    \\ \hline
 \ \  $P_L(H\to\gamma\mu^+\mu^-)$ \ \  & \ \  10.8\% \ \  & \ \  42.3\% \ \  & \ \ 54.1\% \ \  & \ \  -0.17\% \ \  & \ \  -20.0\% \ \  & \ \  -3.3\% \ \    \\ \hline
 \ \  $P_L(H\to\gamma\tau^+\tau^-)$ \ \  & \ \  2.4\% \ \  & \ \  2.6\% \ \  & \ \ 2.0\% \ \  & \ \  1.4\% \ \  & \ \  -1.0\% \ \  & \ \  0.1\% \ \    \\ \hline
 \ \  $P_T(H\to\gamma\mu^+\mu^-)$ \ \  & \ \  -1.0\% \ \  & \ \  -8.2\% \ \  & \ \ 20.1\% \ \  & \ \  -11.2\% \ \  & \ \  2.7\% \ \  & \ \  -0.1\% \ \    \\ \hline
  \ \  $P_T(H\to\gamma e^+e^-)$ \ \  & \ \  -2.0\% \ \  & \ \  -4.0\% \ \  & \ \ -5.0\% \ \  & \ \  -7.4\% \ \  & \ \  1.5\% \ \  & \ \  -1.4\% \ \    \\ \hline
  \ \  $R^{e\mu}$ \ \  & \ \  94.9\% \ \  & \ \  78.8\% \ \  & \ \ 57.8\% \ \  & \ \  84.6\% \ \  & \ \  82.3\% \ \  & \ \ 90.4\% \ \    \\  \hline 
  \ \  $R^{\mu\tau}$ \ \  & \ \  14.1\% \ \  & \ \  3.1\% \ \  & \ \ 1.5\% \ \  & \ \  4.1\% \ \  & \ \  3.5\% \ \  & \ \ 3.6\% \ \    \\  \hline  
 \ \  $R^{e\mu}_{L-}$ \ \  & \ \  94.4\% \ \  & \ \  63.8\% \ \  & \ \ 9.2\% \ \  & \ \  84.8\% \ \  & \ \  85.3\% \ \  & \ \ 90.9\% \ \    \\  \hline
 \ \  $R^{e\mu}_{L+}$ \ \  & \ \  95.4\% \ \  & \ \  84.5\% \ \  & \ \ 72.3\% \ \  & \ \  84.4\% \ \  & \ \  77.9\% \ \  & \ \ 90.5\% \ \    \\  \hline
 \ \  $R^{\mu\tau}_{L-}$ \ \  & \ \  12.8\% \ \  & \ \  1.8\% \ \  & \ \ 0.72\% \ \  & \ \  4.2\% \ \  & \ \  4.2\% \ \  & \ \ 3.86  \% \ \    \\  \hline
  \ \  $R^{\mu\tau}_{L+}$ \ \  & \ \  15.2\% \ \  & \ \  4.4\% \ \  & \ \ 2.3\% \ \  & \ \  4.0\% \ \  & \ \  2.8\% \ \  & \ \ 3.6\% \ \    \\
\hline\hline
\end{tabular}}
\caption{The average polarization asymmetries $\langle
P_{L/T}\rangle$ and ratio of total decay rate $R^{\ell\ell'}_{(i\pm)}$ in $H\to\ell^+\ell^-\gamma$ decays. The uncertainties in these observables is also less than $1\%$.} \label{wc table}
\end{table*}

\subsubsection{Electron polarization asymmetry}
From the three different leptons, lepton polarization asymmetry is most prominent for the case of electrons, approaching $+1$ and $-0.8$ at its extreme values in the plot against $m_{ee}$ which can be seen in Fig. \ref{LET} (a). While the magnitude of extreme values is $-0.1$ $(0.09)$ in the plots against $m_{e\gamma}$. Asymmetry is negligible for normal ($P_N\sim10^{-5}$) and transverse ($P_T\sim10^{-4}$) polarizations in both $m_{ee}$ and $m_{e\gamma}$ plots.

We argue that for electrons, the total decay rate is dominated by loop level contributions which are proportional to $\alpha^{4}$, where $\alpha$ is the QED fine structure constant. Since the physical observable spin asymmetry is a ratio of decay rates, the factors of $\alpha$ almost entirely cancel when calculating polarization asymmetries. Hence, the discrepancies arising from the choice of QED fine structure constant as reported in ref. \cite{Kachanovich:2020xyg} can be ameliorated. Thus, polarization asymmetry may be regarded as a good complementary observable in addition to differential decay rates for the case of final state electrons.

\subsubsection{Muon polarization asymmetry}
In this case the polarization asymmetry is significant for longitudinal as well as transverse polarizations, while remains relatively small for normally polarized case.

\subsubsection{Tauon polarization asymmetry}

In this case the polarization asymmetry remains largely insignificant due to the dominant tree level contribution to the Higgs decay to tauons. 

Finally, we proceed towards the last observable, the ratio of decay rates.

\subsection{Ratio of  differential decay rates $R^{\ell\ell'}$}

\begin{figure}
\centering
\begin{tabular}{cc}
\includegraphics[width=3in,height=2in]{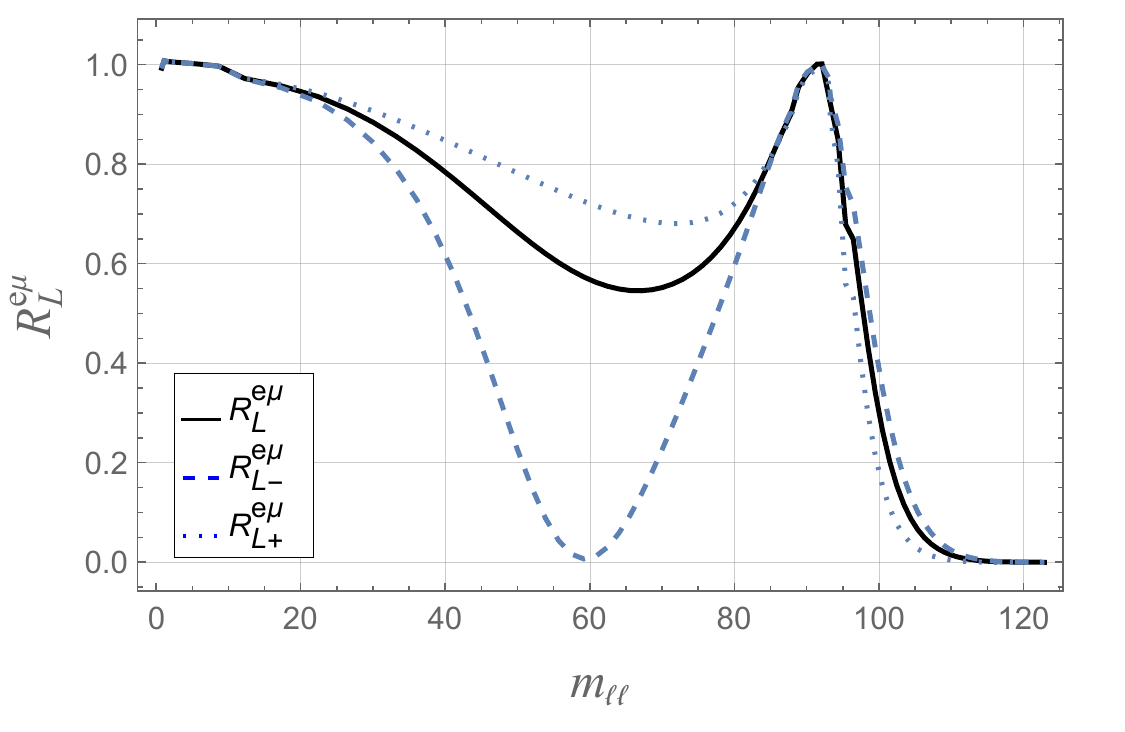}&
\includegraphics[width=3in,height=2in]{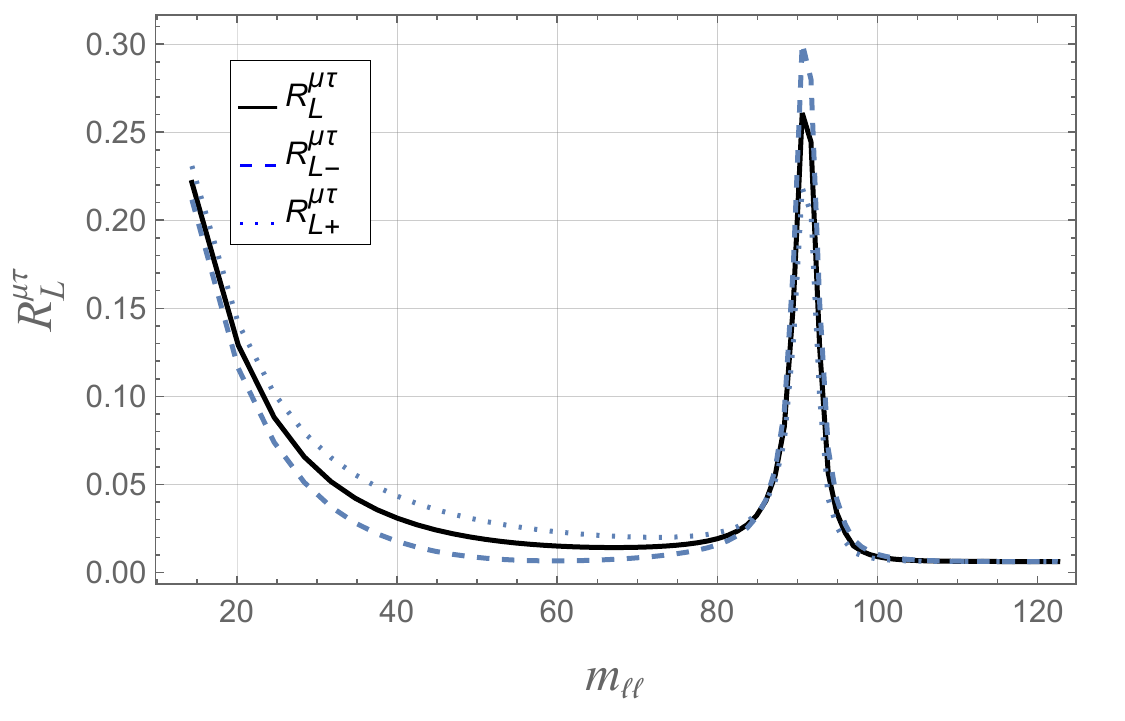}\\
(a)&(b)\\
\includegraphics[width=3in,height=2in]{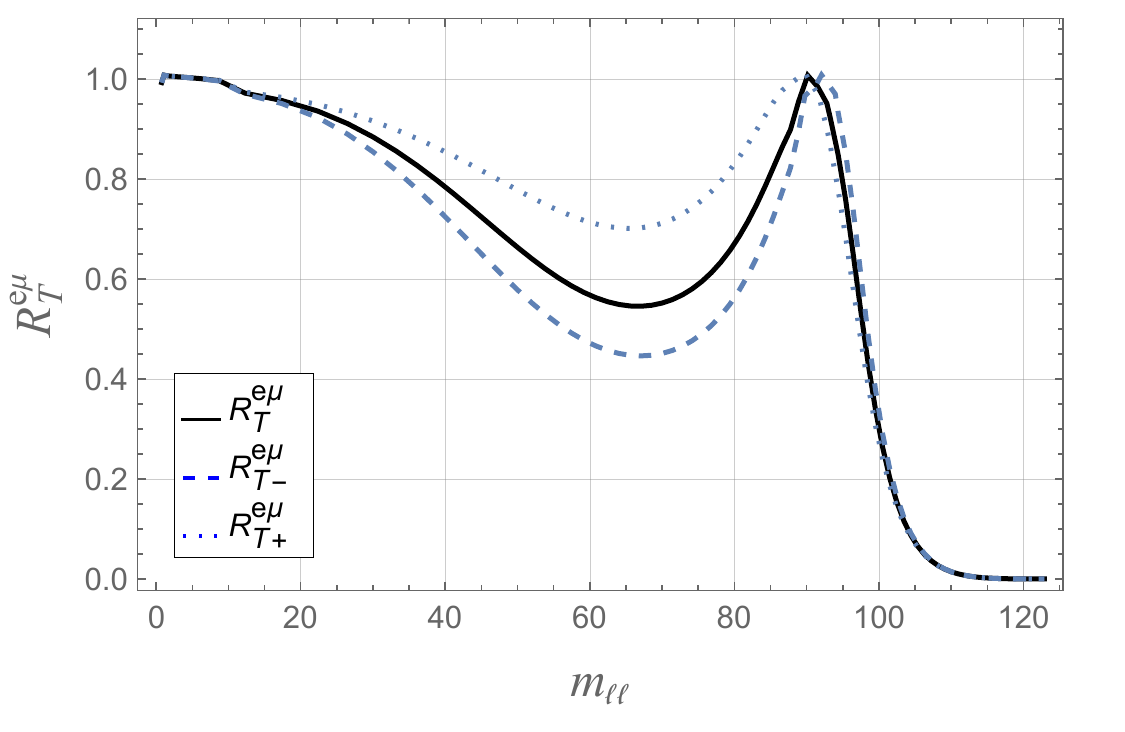}&
\includegraphics[width=3in,height=2in]{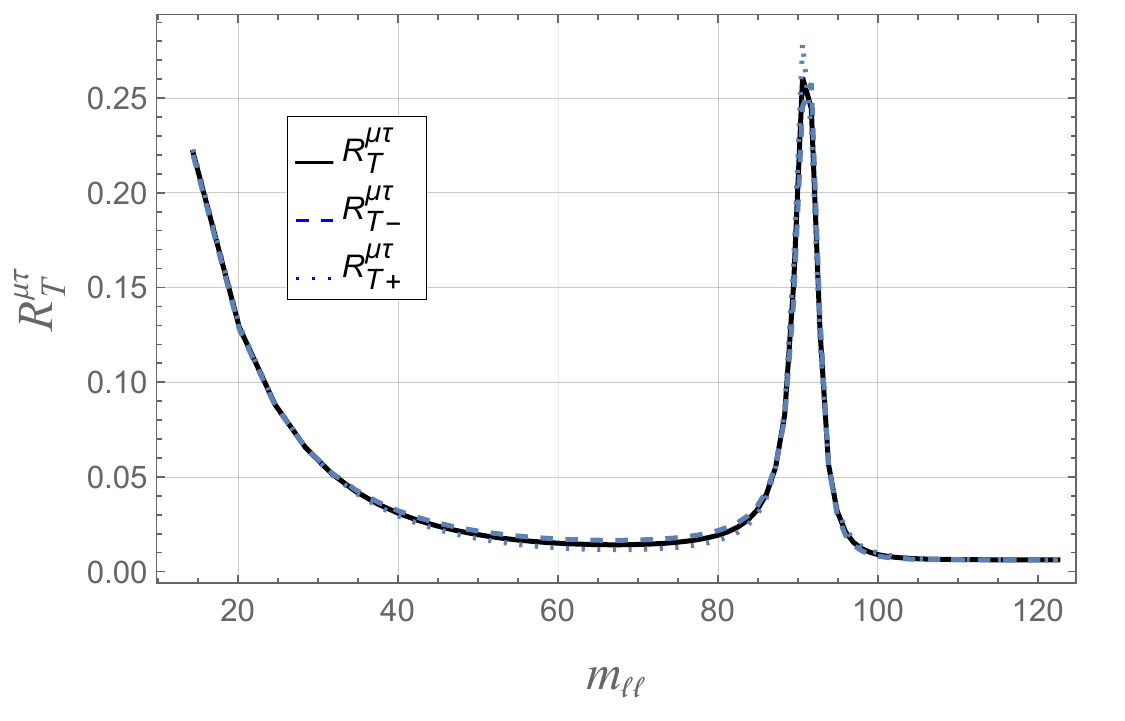}\\
(c)&(d)
\end{tabular}
\caption{Ratio $R_L^{\ell\ell'}$ is plotted in the first row and $R_T^{\ell\ell'}$ in second row as a function of $m_{\ell\ell}$. While the left column represents $R^{e\mu}$ and the right column shows $R^{\mu\tau}$.}
\label{Ratio}
\end{figure}

We will discuss the ratio of decay rates $R^{\ell\ell'}$, given in the expression (\ref{Ratio1}) where $\ell,\ell'$ corresponds to  different generations of dilepton pair in the final state. It is a cleaner observable as common systematic uncertainties cancel out in the numerator and denominator \cite{Bordone:2016gaq,Hurth:2021nsi}. In Fig. \ref{Ratio}, the solid lines show the ratio of unpolarized decay rates, the dotted curves show the ratio of polarized decay rates with spin $+\frac{1}{2}$ and the dashed lines represent the ratio of polarized decay rates with spin $-\frac{1}{2}$.

\subsubsection{Electron-to-muon ratio $R^{e\mu}$}

In Fig. \ref{Ratio}(a,c) we find the ratio $R^{e\mu}$ becomes unity at photon pole and Z pole because in these regions because the loop contribution dominates over the tree contribution. Due to the same reason, the curves exhibiting the ratio of polarized and unpolarized decay rates overlap in this region. In other regions, there is effect of Yukawa coupling which differs for different generations of leptons based on their masses. 

Furthermore, in the region between photon pole and Z pole, the decay rates for the case of final state muon receive considerable tree level contribution. Hence, the decay rate for muon becomes greater than electron, making the  denominator large and the ratio smaller than unity.
Around $60$ GeV region, the ratio $R^{e\mu}_{L-}$ shows a prominent dip, approaching zero. The origin of this dip is the sharp decrease in longitudinally polarized decay rate for final state electron with spin $-\frac{1}{2}$ (numerator) (Fig. \ref{LES}) but comparatively flat behavior in case of final state muon (denominator) (Fig. \ref{LES} e).

After Z pole, the ratio drops to zero because of the tree contribution in muon making denominator much larger as compare to numerator.
For transverse case, the behavior is similar except the magnitude of ratio $R^{e\mu}_{T-}$ is larger as compared to the magnitude of $R^{e\mu}_{L-}$ in the region around $60$ GeV.

\subsubsection{Muon-to-tauon ratio $R^{\mu\tau}$}

This ratio is very small in most of the kinematic region, excluding photon and Z poles, due to the mass dependent Yukawa coupling of tauon in the denominator. The larger ratio near the photon and Z pole is because of the considerably larger loop contribution for the final state muon. In this region, near the photon pole and after the Z pole, the polarized and unpolarized curves coincide. The overlap at photon pole occurs due to absence of polarization effects at low energy. However, after the Z pole, overlap occurs due to the tree dominance in both numerator $d\Gamma(m_{\mu\mu})$ and denominator $d\Gamma(m_{\tau\tau})$. Furthermore, in the region around $60$ GeV, the three curves split especially for longitudinally polarized case, following the pattern of muon (appearing in numerator), as the denominator stays the same for polarized and unpolarized cases.

\section{Conclusion}\label{con}
We have performed a complete calculation of lepton spin-polarized decay rates including all contributions up to one-loop level for $H\rightarrow \ell^{+}\ell^{-}\gamma$. For this process, we have also presented our results for polarization asymmetries, as well as the ratio of decay rates for different final state lepton generations. We \textit{keep track} of resonant and non-resonant contributions in polarized decay rates which include the contribution from the on-shell decay $H\rightarrow Z[\rightarrow \ell^{+}\ell^{-}]\gamma$.  We observe that for $\ell=e,\mu$, the longitudinal decay rate is highly suppressed around $m_{\ell\ell}\approx 60$ GeV when the final lepton spin is $-\frac{1}{2}$, dramatically increasing the corresponding lepton polarization asymmetries. This effect gets progressively smaller with increasing lepton mass, i.e., it is less noticeable for final state muons, and nearly negligible for tauons. The same phenomenon is also observed in longitudinal and transverse electron-to-muon ratio $R^{e\mu}$ that shows a dip at $60-70$ GeV. Therefore, we propose that final state lepton polarization dependent observables, such as polarization asymmetries can serve as an important probe for investigating properties of the Higgs particle. Precise measurements of these observables at the HL-LHC and the planned $e^{+}e^{-}$ can provide a fertile ground to test not only the SM but also to examine the signatures of possible NP beyond the SM. 

\begin{appendix}

\section{Appendix}
In this Appendix, formulas for $a_{1,2}$ and $b_{1,2}$ are given,  which are introduced in Eq. \ref{Cfunctions} and characterize the decay rate given in Eq. \ref{ML}. We have applied $D=4-2\epsilon$ which appears in the coefficients of function $B_0$. When $B_0$ is expanded in $\epsilon$ to order $\epsilon^0$, the pole $1/\epsilon$ vanishes from $a_1$ and $b_1$. The function $a_1$ reads:
{\footnotesize{
\begin{eqnarray}
a_1 &=&\frac{e^4}{48 \pi ^2
   x^3 \sqrt{1-x^2} m_Z}\bigg[\frac{1}{\mathcal{B}
   \left(m_H^2-s\right)^2}\{m_H^2\{-3 \left(2
   x^2-1\right) s (B_0^x-B_0^{sx}-1)+4
   \left(8 x^2-5\right) m_t^2\notag \\
   &+&6 x^2
   \left(1-6 x^2\right) m_Z^2\}-6 x^2
   \left(6 x^2-1\right) m_Z^2 s
   (B_0^x-B_0^{sx}-1)+4 \left(8 x^2-5\right)
   m_t^2 s (B_0^{m_t}-B_0^{s}-1)\notag \\
   &+&\left(3-6
   x^2\right) m_H^4\}+\frac{2 x^2
   \left(18 x^2 m_Z^2 (B_0^x-B_0^{sx}-1)+3
   m_H^2 (B_0^x-B_0^{sx}-1)+16 m_t^2
   (-B_0^{m_t}+B_0^{s}+1)\right)}{\left(m_H^2-s\right)^2}\notag \\
   &+&\frac{6 \left(x^2-1\right) m_Z^2 \left(-2
   B_0^{m_Z}+2 B_0^{u}-C_0^{u}
   \left(m_H^2+m_Z^2-s\right)+m_Z^2
   (C_0^{uZ}+C_0^{t})-C_0^{uZ} s+C_0^{tz}
   \left(m_H^2+m_Z^2\right)\right)}{s t}\notag \\
   &-&\frac{12
   \left(x^2-1\right) m_Z^2
   \left(-B_0^{m_Z}+B_0^{u}+C_0^{uZ}
   m_Z^2\right)}{s (s+t)}+\frac{2 x^2
   m_H^2 \left(18 x^2 m_Z^2+3 m_H^2-16
   m_t^2\right)}{s
   \left(m_H^2-s\right)^2}\notag \\
   &-&\frac{2
   \left(C_0^{st} \left(8 x^2-5\right) m_t^2
   \left(m_H^2-4 m_t^2-s\right)+3 C_0^{sx}
   x^2 m_Z^2 \left(\left(6 x^2-1\right)
   \left(2 x^2 m_Z^2-m_H^2\right)+8
   x^2 s-2 s\right)\right)}{\mathcal{B}
   \left(m_H^2-s\right)}\notag \\
   &+&\frac{16 x^2 \left(3
   C_0^{sx} x^2 m_Z^2-C_0^{st}
   m_t^2\right)}{m_H^2-s}+\frac{4 x^2
   \left(4 C_0^{st} m_t^2 \left(m_H^2-4
   m_t^2\right)+18 C_0^{sx} x^4 m_Z^4-9
   C_0^{sx} x^2 m_H^2 m_Z^2\right)}{s
   \left(m_H^2-s\right)}\notag \\
   &+&\frac{6
   \left(x^2-1\right) m_Z^2
   (C_0^{u}+C_0^{uZ}-C_0^{t}-C_0^{tz})}{s}+\frac
   {6 \left(x^2-1\right) m_Z^4 \left(C_0^{u}
   \left(m_H^2-s\right)+C_0^{uZ}
   s-C_0^{tz} m_H^2\right)}{s t^2}\notag \\
   &+&\frac{6
   \left(x^2-1\right) D_0^{u} m_Z^2
   \left(m_Z^2-t\right) \left(m_H^2
   \left(m_Z^2-t\right)-m_Z^2 s+t
   (s+t)\right)}{s t^2}\bigg].
\end{eqnarray}
}}
The explicit form of the function $a_2$ is:
{\footnotesize{
\begin{eqnarray}
   a_2&=& \frac{e^4} {48 x^3\sqrt{1-x^2}m_Z \pi^2}\bigg[\frac{12 C_0^{sx} m_Z^2 \left(-3
   m_H^2+6 x^2 m_Z^2+4 s\right)
   x^4}{\left(m_H^2-s\right) s}-\frac{16
   C_0^{st} m_t^2 \left(-m_H^2+4
   m_t^2+s\right)
   x^2}{\left(m_H^2-s\right) s}\notag \\
   &-&\frac{6 C_0^{sx} m_Z^2 \left(8 s
   x^2+\left(6 x^2-1\right) \left(2 x^2
   m_Z^2-m_H^2\right)-2 s\right)
   x^2}{\mathcal{B}
   \left(m_H^2-s\right)}-\frac{32
   m_t^2 \left(m_H^2+B_0^{m_t} s\right)
   x^2}{\left(m_H^2-s\right)^2
   s}+\frac{32 (B_0^{s}+1) m_t^2
   x^2}{\left(m_H^2-s\right)^2}\notag \\
   &+&\frac{6 \left(m_H^4+\left(6
   x^2 m_Z^2+B_0^{x} s\right)
   m_H^2+2 (3 B_0^{x}-2) x^2 m_Z^2
   s\right) x^2}{\left(m_H^2-s\right)^2
   s}-\frac{6
   \left((B_0^{sx}+1) m_H^2+2 (3 B_0^{sx}+1) x^2
   m_Z^2\right)
   x^2}{\left(m_H^2-s\right)^2}\notag \\
   &-&\frac{4
   (B_0^{s}+1) \left(8 x^2-5\right) m_t^2
   s}{\mathcal{B} \left(m_H^2-s\right)^2}+\frac{2
   C_0^{st} \left(8 x^2-5\right) m_t^2
   \left(-m_H^2+4 m_t^2+s\right)}{\mathcal{B}
   \left(m_H^2-s\right)}+\frac{4 \left(8
   x^2-5\right) m_t^2 \left(m_H^2+B_0^{m_t}
   s\right)}{\mathcal{B}
   \left(m_H^2-s\right)^2}\notag \\
   &+&\frac{3
   \left((B_0^{sx}+1) \left(2 x^2-1\right) m_H^2+2
   x^2 \left(2 (3 B_0^{sx}+1)
   x^2-B_0^{sx}-1\right) m_Z^2\right)
   s}{\mathcal{B} \left(m_H^2-s\right)^2}-\frac{12 (B_0^{t}-B_0^{m_z})
   \left(x^2-1\right) m_Z^2}{\left(m_H^2-t\right)
   u}\notag \\
   \notag \\
   &+&\frac{3}{\mathcal{B}
   \left(m_H^2-s\right)^2} \{\left(1-2
   x^2\right) m_H^4+\left(2 \left(\left(1-6
   x^2\right) m_Z^2-B_0^{x} s\right)
   x^2+B_0^{x} s\right) m_H^2 +2
   x^2 (-6 B_0^{x} x^2+4
   x^2+B_0^{x}) m_Z^2
   s\}\notag \\
   &-&\frac{6 C_0^{u}
   \left(x^2-1\right) m_Z^2
   \left(m_Z^2-u\right)}{s
   \left(-u\right)}-\frac{6 C_0^{uZ}
   \left(x^2-1\right) m_Z^2 (s+t)
   \left(m_Z^2-u\right)}{s
   \left(-u\right)^2}+\frac{6 C_0^{t}
   \left(x^2-1\right) m_Z^2 t
   \left(m_Z^2-u\right)}{s
   \left(-u\right)^2}\notag \\
    &+&\frac{6
   \left(x^2-1\right) D_0^{tu} m_Z^2
   \left(m_Z^2-u\right)
   \left(\left(m_Z^2-t\right) m_H^2-m_Z^2 s+t
   (s+t)\right)}{s
   \left(-u\right)^2}-\frac{6 C_0^{tz}
   \left(x^2-1\right) m_Z^2}{s \left(m_H^2-t\right)
   \left(-u\right)^2}\{-m_H^6+(m_Z^2\notag
      \end{eqnarray}
   \begin{eqnarray}
     &+&3 (s+t))
   m_H^4-\left(2 (2 s+t) m_Z^2+2 s^2+3 t^2+6
   s t\right) m_H^2+t (s+t) (2
   s+t)+m_Z^2 \left(2 s^2+4 t
   s+t^2\right)\}\bigg].
\end{eqnarray}
}}
The explicit form of the function $b_1$ is:
{\footnotesize{
\begin{eqnarray}
   b_1&=& \frac{\text{e}^4}{96 x^3 \left(1-x^2\right)^{3/2}
   m_Z \pi ^2}\bigg[-12 D_0^{tx} m_Z^2
   x^4-\frac{6 C_0^{ux} m_Z^2
   \left(m_H^2-s-t\right) \left(-x^2
   m_Z^2+s+t\right) x^4}{s t^2}\notag \\
   &-&\frac{6
   C_0^{uH} m_Z^2 \left(x^2
   m_Z^2-s-t\right) \left(s^2+2 t
   s-t^2\right) x^4}{s t^2
   (s+t)}+\frac{6 D_0^t m_Z^2 \left(x^4
   \left(m_H^2-s\right) m_Z^4-x^2
   \left(m_H^2-2 s\right) t m_Z^2+s
   t^2\right) x^4}{s t^2}\notag \\
   &+&\frac{6 C_0^{sx}
   m_Z^2 \left(\left(s-2 x^2 m_Z^2\right)
   \left(m_H^2-s\right)^2+2 t
   \left(m_H^2-s\right)^2-4 \left(x^2-1\right)
   \left(-3 m_H^2+6 x^2 m_Z^2+4 s\right)
   t^2\right) x^4}{\left(m_H^2-s\right)
   s t^2}\notag \\
   &-&\frac{6 D_0^{ux} m_Z^2 \left(x^4
   \left(s-m_H^2\right) m_Z^4+x^2
   \left(m_H^2 (2 s+t)-2 s (s+2 t)\right)
   m_Z^2+s (s+t)
   \left(-m_H^2+s+t\right)\right) x^4}{s
   t^2}\notag \\
   &+&\frac{6 C_0^{tx} m_Z^2 \left(x^2
   m_Z^2-t\right) x^4}{s t}-\frac{12
   (B_0^{ux}-1) m_Z^2 x^4}{t (s+t)}-\frac{6
   C_0^{sz} m_Z^2 s x^4}{t^2}+\frac{6
   C_0^{tH} m_Z^2 \left(m_H^2-t\right)
   \left(x^2 m_Z^2-t\right) x^4}{s
   t^2}\notag \\
   &+&\frac{32 C_0^{st} \left(x^2-1\right) m_t^2
   \left(-m_H^2+4 m_t^2+s\right)
   x^2}{\left(m_H^2-s\right) s}+\frac{64
   \left(x^2-1\right) m_t^2
   \left(m_H^2+B_0^{m_t} s\right)
   x^2}{\left(m_H^2-s\right)^2
   s}\notag \\
   &+&\frac{6 C_0^{sx} \left(2 x^2-1\right)
   m_Z^2 \left(8 s x^2+\left(6
   x^2-1\right) \left(2 x^2
   m_Z^2-m_H^2\right)-2 s\right)
   x^2}{\mathcal{B}
   \left(m_H^2-s\right)} +\frac{4
   (B_0^{s}+1) \left(16 x^4-18 x^2+5\right)
   m_t^2 s}{\mathcal{B}
   \left(m_H^2-s\right)^2}\notag \\
   &+&\frac{12
   \left(\frac{(B_0^x-1) x^2
   m_Z^2}{t}-\frac{(B_0^x-1) x^2
   m_Z^2}{s+t}-\frac{\left(x^2-1\right)
   \left(m_H^4+\left(6 x^2 m_Z^2+B_0^x
   s\right) m_H^2+2 (3 B_0^x-2) x^2
   m_Z^2
   s\right)}{\left(m_H^2-s\right)^2}\right)
   x^2}{s}\notag \\
   &-&\frac{64 (B_0^{s}+1)
   \left(x^2-1\right) m_t^2
   x^2}{\left(m_H^2-s\right)^2}+\frac{12
   \left(x^2-1\right) \left((B_0^{sx}+1) m_H^2+2 (3
   B_0^{sx}+1) x^2 m_Z^2\right)
   x^2}{\left(m_H^2-s\right)^2}\notag \\
   &-&\frac{3 \left(2
   x^2-1\right) \left((B_0^{sx}+1) \left(2
   x^2-1\right) m_H^2+2 x^2 \left(2 (3
   B_0^{sx}+1) x^2-B_0^{sx}-1\right)
   m_Z^2\right) s}{\mathcal{B}
   \left(m_H^2-s\right)^2}\notag \\
   &-&\frac{2 C_0^{st} \left(16
   x^4-18 x^2+5\right) m_t^2
   \left(-m_H^2+4 m_t^2+s\right)}{\mathcal{B}
   \left(m_H^2-s\right)} -\frac{4 \left(16 x^4-18
   x^2+5\right) m_t^2 \left(m_H^2+B_0^{m_t}
   s\right)}{\mathcal{B}
   \left(m_H^2-s\right)^2}\notag \\
  &+&\frac{3 \left(2
   x^2-1\right)}{\mathcal{B}
   \left(m_H^2-s\right)^2} \{\left(2 x^2-1\right)
   m_H^4+(2 x^2 \left(\left(6
   x^2-1\right) m_Z^2+B_0^x
   s\right)-B_0^x s) m_H^2+2
   x^2 \left(2 (3 B_0^x-2)
   x^2-B_0^x\right) m_Z^2
   s\}\notag \\
   &-&\frac{3 C_0^u \left(1-2
   x^2\right)^2 m_Z^2 \left(m_Z^2-t\right)
   \left(u\right)}{s t^2}-\frac{3
   D_0^u \left(m_Z-2 x^2 m_Z\right)^2
   \left(m_Z^2-t\right) \left(\left(m_Z^2-t\right)
   m_H^2-m_Z^2 s+t (s+t)\right)}{s
   t^2}\notag \\
     &-&\frac{3 C_0^u \left(m_Z-2 x^2
   m_Z\right)^2 \left(t^3-s^2 t+m_Z^2
   \left(s^2+2 t s-t^2\right)\right)}{s t^2
   (s+t)}-\frac{3 C_0^t \left(1-2 x^2\right)^2
   m_Z^2 \left(m_Z^2-t\right)}{s t}\notag \\
   &+&\frac{6
   (B_0^{m_Z}-1) \left(1-2 x^2\right)^2 m_Z^2}{t
   (s+t)}-\frac{6 (B_0^{u}-1) \left(1-2
   x^2\right)^2 m_Z^2}{t (s+t)}+\frac{3
   C_0^{tz} \left(1-2 x^2\right)^2 m_Z^2
   \left(m_H^2-t\right) \left(m_Z^2-t\right)}{s
   t^2}\bigg].
\end{eqnarray}
}}
The explicit form of the function $b_2$ is:
{\footnotesize{
\begin{eqnarray}
 b_2&=&\frac{e^4}{96 x^3
   \left(1-x^2\right)^{3/2} m_Z \pi ^2}\bigg[-12 D_0^{su} m_Z^2
   x^4+\frac{6 C_0^{uH} m_Z^2 (s+t)
   \left(x^2 m_Z^2-u\right)
   x^4}{s
   \left(-u\right)^2}+ \frac{6 C_0^{ux}
   m_Z^2 \left(-\frac{x^2
   m_Z^2}{-u}-1\right)
   x^4}{s}\notag \\
   &+&+\frac{6 C_0^{sx} \left(2 x^2-1\right)
   m_Z^2 \left(8 s x^2+\left(6
   x^2-1\right) \left(2 x^2
   m_Z^2-m_H^2\right)-2 s\right)
   x^2}{\mathcal{B}
   \left(m_H^2-s\right)}\notag \\
   &+&\frac{6 C_0^{sx} m_Z^2
   \left(\frac{2
   \left(m_H^2-s\right)^2}{-u}-\frac
   {\left(s-2 x^2 m_Z^2\right)
   \left(m_H^2-s\right)^2}{\left(-u\right)^2}+4 \left(x^2-1\right) \left(-3 m_H^2+6
   x^2 m_Z^2+4 s\right)\right)
   x^4}{s \left(s-m_H^2\right)}\notag \\
   &+&\frac{6
   D_0^{ux} m_Z^2 \left(x^4
   \left(m_H^2-s\right) m_Z^4-x^2
   \left(m_H^2-2 s\right)
   \left(m_H^2-s-t\right) m_Z^2+s
   \left(-u\right)^2\right)
   x^4}{s
   \left(-m_H^2+s+t\right)^2}\notag \\
   &+&\frac{6 C_0^{tH}
   m_Z^2 \left(-m_H^2+x^2 m_Z^2+t\right)
   \left(m_H^4-2 (2 s+t) m_H^2+2 s^2+t^2+4
   s t\right) x^4}{s
   \left(m_H^2-t\right)
   \left(-u\right)^2}\notag \\
   &+&\frac{6 D_0^{t}
   m_Z^2 \left(x^4 \left(m_H^2-s\right)
   m_Z^4-x^2 \left(m_H^4-(3 s+t)
   m_H^2+2 s (s+2 t)\right) m_Z^2+s
   \left(m_H^2-t\right) t\right) x^4}{s
   \left(-u\right)^2}\notag \\
   &-&\frac{12 (B_0^{tu}-1)
   m_Z^2 x^4}{\left(m_H^2-t\right)
   \left(u\right)}-\frac{6 C_0^{sz} m_Z^2
   s
   x^4}{\left(-u\right)^2}
   +\frac{6
   C_0^{tx} m_Z^2 t \left(-m_H^2+x^2
   m_Z^2+t\right) x^4}{s
   \left(-u\right)^2}\notag \\
   &+&\frac{32 C_0^{st}
   \left(x^2-1\right) m_t^2 \left(-m_H^2+4
   m_t^2+s\right)
   x^2}{\left(m_H^2-s\right) s}+\frac{64
   \left(x^2-1\right) m_t^2
   \left(m_H^2+B_0^{m_t} s\right)
   x^2}{\left(m_H^2-s\right)^2
   s}\notag \\
   &+&\frac{12
   \left(\frac{(B_0^{x}-1) x^2
   m_Z^2}{t-m_H^2}-\frac{(B_0^{x}-1) x^2
   m_Z^2}{-u}-\frac{\left(x^2-1\right) \left(m_H^4+\left(6 x^2
   m_Z^2+B_0^{x} s\right) m_H^2+2 (3
   B_0^{x}-2) x^2 m_Z^2s\right)}{\left(m_H^2-s\right)^2}\right)
   x^2}{s}\notag \\
   &-&\frac{64 (B_0^{s}+1)
   \left(x^2-1\right) m_t^2
   x^2}{\left(m_H^2-s\right)^2}+\frac{12
   \left(x^2-1\right) \left((B_0^{sx}+1) m_H^2+2 (3
   B_0^{sx}+1) x^2 m_Z^2\right)
   x^2}{\left(m_H^2-s\right)^2}\notag \\
   &+&\frac{4
   (B_0^{s}+1) \left(16 x^4-18 x^2+5\right)
   m_t^2 s}{\mathcal{B}
   \left(m_H^2-s\right)^2}\notag \\
 &-&\frac{3 \left(2
   x^2-1\right) \left((B_0^{sx}+1) \left(2
   x^2-1\right) m_H^2+2 x^2 \left(2 (3
   B_0^{sx}+1) x^2-B_0^{sx}-1\right)
   m_Z^2\right) s}{\mathcal{B}
   \left(m_H^2-s\right)^2}\notag \\
     &-&\frac{2 C_0^{st} \left(16
   x^4-18 x^2+5\right) m_t^2
   \left(-m_H^2+4 m_t^2+s\right)}{\mathcal{B}
   \left(m_H^2-s\right)}-\frac{4 \left(16 x^4-18
   x^2+5\right) m_t^2 \left(m_H^2+B_0^{m_t}
   s\right)}{\mathcal{B}
   \left(m_H^2-s\right)^2}   
   \notag \\
   &+&\frac{3 \left(2
   x^2-1\right)}{\mathcal{B}
   \left(m_H^2-s\right)^2} \{\left(2 x^2-1\right)
   m_H^4+\left(2 x^2 \left(\left(6
   x^2-1\right) m_Z^2+B_0^{x}
   s\right)-B_0^{x} s\right) m_H^2\notag \\
    &+&2
   x^2 \left(2 (3 B_0^{x}-2)
   x^2-B_0^{x}\right) m_Z^2
   s\}+\frac{3 C_0^{u} \left(1-2
   x^2\right)^2 m_Z^2
   \left(m_Z^2-u\right)}{s
   \left(-u\right)}\notag \\
   &-&\frac{3 C_0^{t} \left(1-2
   x^2\right)^2 m_Z^2 t
   \left(m_Z^2-u\right)}{s
   \left(-u\right)^2}+\frac{3 C_0^{uZ}
   \left(1-2 x^2\right)^2 m_Z^2 (s+t)
   \left(m_Z^2-u\right)}{s
   \left(-u\right)^2}\notag \\
  &-&\frac{3 D_0^{tu}
   \left(m_Z-2 x^2 m_Z\right)^2
   \left(m_Z^2-u\right)
   \left(\left(m_Z^2-t\right) m_H^2-m_Z^2 s+t
   (s+t)\right)}{s
   \left(-u\right)^2}\notag \\
   &+&\frac{3 C_0^{tz}
   \left(m_Z-2 x^2 m_Z\right)^2}{s \left(m_H^2-t\right)
   \left(-u\right)^2}
   \{-m_H^6+\left(m_Z^2+3 (s+t)\right)
   m_H^4-\left(2 (2 s+t) m_Z^2+2 s^2+3 t^2+6
   s t\right) m_H^2\notag \\
   &+&t (s+t) (2
   s+t)+m_Z^2 \left(2 s^2+4 t
   s+t^2\right)\}+\frac{6 (B_0^{m_Z}-B_0^{u})
   \left(1-2 x^2\right)^2
   m_Z^2}{\left(m_H^2-t\right)
   \left(u\right)}\bigg].
\end{eqnarray}
}}
where $\mathcal{B}=s-m_Z^2+im_Z\Gamma_Z$, $x\equiv C_W$ and $B_0^i$, $C_0^i$ and $D_0^i$ are given in terms of reduced Passarino-Veltman functions as follows,
{\footnotesize{
\begin{eqnarray}
B_0^{m_t}&\equiv&\widetilde{\text{B}}_0[m_H^2,m_t^2,m_t^2];\qquad\qquad B_0^{m_Z}\equiv\widetilde{\text{B}}_0[m_H^2,m_Z^2,m_Z^2]\qquad\qquad B_0^{u}\equiv\widetilde{\text{B}}_0[u,0,m_Z^2]\notag\\
B_0^{x}&\equiv&\widetilde{\text{B}}_0[m_H^2,x^2 m_Z^2,x^2
   m_Z^2];\qquad\;\; B_0^{s}\equiv\widetilde{\text{B}}_0[s,m_t^2,m_t^2];\qquad\qquad\;\;
    B_0^{sx}\equiv\widetilde{\text{B}}_0[s,x^2 m_Z^2,x^2
   m_Z^2]\notag \\     B_0^{ux}&\equiv&\widetilde{\text{B}}_0[u,0,x^2 m_Z^2];\quad\qquad\qquad B_0^{tu}\equiv \widetilde{\text{B}}_0[t, 0, x^2*m_Z^2]
   \qquad\qquad
C_0^{uZ}\equiv\text{C}_0[0,m_H^2,u,0,m_Z^2,m_Z^2]\notag \\
C_0^t&\equiv&\text{C}_0[0,0,t,0,0,m_Z^2];\qquad
   C_0^{tz} \equiv\text{C}_0[0,m_H^2,t,0,m_Z^2,m_Z^2];\qquad
C_0^{st}\equiv\text{C}_0[0,m_H^2,s,m_t^2,m_t^2,m_t^2]\notag \\
   C_0^{sx} &\equiv&\text{C}_0[0,m_H^2,s,x^2 m_Z^2,x^2
   m_Z^2,x^2 m_Z^2];\qquad\qquad
  C_0^{sz}\equiv\text{C}_0[0,0,s,x^2 m_Z^2,0,x^2
   m_Z^2]\notag \\
  C_0^{ux} &\equiv&\text{C}_0[0,0,u,0,x^2
   m_Z^2,x^2 m_Z^2];\qquad\qquad\qquad\quad
  C_0^{tx} \equiv\text{C}_0[0,0,t,0,x^2 m_Z^2,x^2
   m_Z^2]\notag \\
  C_0^{uH} &\equiv&\text{C}_0[0,m_H^2,u,0,x^2
   m_Z^2,x^2 m_Z^2];\qquad\qquad\qquad
  C_0^{tH}\equiv\text{C}_0[0,m_H^2,t,0,x^2
   m_Z^2,x^2 m_Z^2]\notag \\
D_0^u&\equiv&\text{D}_0[0,u,0,t,0,m_H^2,0,m_Z^2,0,m_Z^2]\qquad\qquad C_0^u \equiv\text{C}_0[0,0,u,0,0,m_Z^2];\notag \\
   D_0^{ux}&\equiv&\text{D}_0[0,u,m_H^2,s,0,0,x^2 m_Z^2,0,x^2
   m_Z^2,x^2 m_Z^2]\notag \\
   D_0^t&\equiv&\text{D}_0[0,0,t,m_H^2,s,0,0,x
   ^2 m_Z^2,0,x^2 m_Z^2,x^2
   m_Z^2]\notag \\
   D_0^{tx}&\equiv&\text{D}_0[0,m_H^2,0,0,0,s,t,x
   ^2 m_Z^2,x^2 m_Z^2,x^2
   m_Z^2,0]\notag \\
   D_0^{su}&\equiv& \text{D}_0[m_H^2,0,0,0,s,u,x^2
   m_Z^2,x^2 m_Z^2,x^2
   m_Z^2,0]\notag \\
  D_0^{tu} &\equiv&\text{D}_0[0, t, 0, u, 0, m_H^2, 0, m_Z^2, 0, m_Z^2].
\end{eqnarray}
}}
Here tilde ( $\widetilde{}$ ) means no $1/\epsilon$ UV-divergent pole in $B_0$. 

\section{Appendix}
The loop contributions arising in Eq. \ref{2.10}, $\vert\mathcal{M}^{i^\pm}_{loop}(s,t)\vert^2$ and the loop-tree interference terms $\vert\mathcal{M}^{i\pm}_{tree-loop}\vert^2$ read as follows:

{\footnotesize{
\begin{eqnarray}
&&\vert\mathcal{M}^{L\pm}_{loop}\vert^2=\frac{s}{2}\bigg[(\vert \mathcal{C}_1\vert^2+\vert \mathcal{C}_2\vert^2)t^2+(\vert \mathcal{C}_3\vert^2+\vert \mathcal{C}_4\vert^2\pm2\text{Re}[\mathcal{C}_3\mathcal{C}_4^*]v)u^2\bigg]\notag \\
&&\mp\frac{st}{2}\text{Re}[\mathcal{C}_1\mathcal{C}_2^*]  \left(
   (s-m_H^2)v+x \sqrt{\lambda  }\right)\pm\frac{s}{4}\left(
   (s-m_H^2+2t)v+x \sqrt{\lambda  }\right)\bigg[\text{Re}[\mathcal{C}_1\mathcal{C}_4^*] (t+u)-\text{Re}[\mathcal{C}_2\mathcal{C}_3^*] (t-u)\bigg] \notag \\
   &&+m_{\ell}^2\bigg[-\vert \mathcal{C}_1\vert^2st-\vert \mathcal{C}_3\vert^2su-\vert \mathcal{C}_2\vert^2t(s+2t)-\vert \mathcal{C}_4\vert^2u(s+2u)\pm\frac{\text{Re}[\mathcal{C}_1\mathcal{C}_2^*] (s+2 (t+u)) \left(v
   \left(s-m_H^2\right)+x \sqrt{\lambda}\right)}{2}\notag \\
   &&+\text{Re}[\mathcal{C}_1\mathcal{C}_3^*](t+u)^2\mp\frac{\text{Re}[\mathcal{C}_1\mathcal{C}_4^*]  \left(-m_H^2v
   (s+2 (t+u))+s^2 v+s
   \left(v (5 t+3 u)+x \sqrt{\lambda }\right)+2 x
   \sqrt{\lambda } (t+u)\right)}{2}\notag \\
&&\pm\frac{1}{2} \text{Re}[\mathcal{C}_2\mathcal{C}_3^*] sv (t-u)+ \text{Re}[\mathcal{C}_2\mathcal{C}_4^*](t-u)^2\mp2\text{Re}[\mathcal{C}_3\mathcal{C}_4^*] usv\bigg]+\frac{m_{\ell}^4}{2}\bigg[s\left(\vert \mathcal{C}_1\vert^2+\vert \mathcal{C}_3\vert^2\right)\notag \\
  &&+\left(\vert \mathcal{C}_2\vert^2+\vert \mathcal{C}_4\vert^2\right)(s+8t)\mp4\text{Re}[\mathcal{C}_1\mathcal{C}_2^*] \left(v
   \left(s-m_H^2\right)+x \sqrt{\lambda 
  }\right)-8\text{Re}[\mathcal{C}_1\mathcal{C}_3^*](t+u)\notag \\
  &&\pm2\text{Re}[\mathcal{C}_1\mathcal{C}_4^*] \left(v
   \left(3s-2m_H^2\right)+2x \sqrt{\lambda 
  }\right)\pm2\text{Re}[\mathcal{C}_3\mathcal{C}_4^*]sv\bigg]-2m_{\ell}^6\left(\vert \mathcal{C}_2\vert^2+\vert \mathcal{C}_4\vert^2-2\text{Re}[\mathcal{C}_1\mathcal{C}_3^*]\right).\label{M21}
\end{eqnarray}

\begin{eqnarray}
&& \vert\mathcal{M}^{L\pm}_{tree-loop}\vert^2=\frac{4 m_\ell}{\left(m_{\ell}^2-t\right)\left(m_{\ell}^2-u\right)}\bigg[ \text{Re}[\mathcal{C}_0\mathcal{C}_1^*]  (12 m_{\ell}^6-2 m_{\ell}^4 (s+8 t+6 u)+m_{\ell}^2 (t+u) (2s+7 t+3 u)\notag \\
   &&-t \left(2 s u+(t+u)^2\right))+ \text{Re}[\mathcal{C}_0\mathcal{C}_3^*]  (12 m_{\ell}^6-2 m_{\ell}^4 (s+6 t+8 u)+m_{\ell}^2 (t+u) (2s+3 t+7 u)\notag \\
   &&-u \left(2s t+(t+u)^2\right))\pm\frac{1}{ m }\text{Re}[\mathcal{C}_0\mathcal{C}_2^*]\bigg[ 8 m_{\ell}^6 \left(2 m_H^2 v-3 sv-2 x \sqrt{\lambda }\right)\notag \\
&&+2 m_{\ell}^4 \left(-2 m_H^2 v
   (2 s+t+3 u)+5 s^2 v+2 s \left(2 x \sqrt{\lambda}+5 t v+3 uv\right)+2 x \sqrt{\lambda s} (t+3u)\right)\notag \\
   &&+m_{\ell}^2 \bigg[m_H^2v \left(s^2+6 s t+2s u-2 t^2+2 u^2\right)-s^3 v-s^2 \left(x\sqrt{\lambda }+2v (5 t+u)\right)\notag \\
   &&-2 s \left(3 t x\sqrt{\lambda }+u x \sqrt{\lambda }+3 t^2 v+u^2v\right)+2 x \sqrt{\lambda  } (t^2-u^2)\bigg]+s^2 t\left(v \left(-m_H^2+s+2 t\right)+x \sqrt{\lambda }\right)\notag \\
   &&\mp\frac{1}{ m }\text{Re}[\mathcal{C}_0\mathcal{C}_4^*]\bigg[ (12 m_{\ell}^6 \left(2 m_H^2 v-3 sv-2 x \sqrt{\lambda }\right)\notag \\
&&+2 m_{\ell}^4 \left(-m_H^2 v
   (5 s+6t+10 u)+6 s^2 v+ s \left(5 x \sqrt{\lambda}+13 t v+17 uv\right)+2 x \sqrt{\lambda s} (3t+5u)\right)\notag \\
   &&+m_{\ell}^2 \bigg[m_H^2v \left(s^2+3 s (t+3u)+4 u(t+u)\right)-s^3 v-s^2 \left(x\sqrt{\lambda }+5 tv+11uv)\right)\notag \\
   &&- s \left(3 t x\sqrt{\lambda }+9u x \sqrt{\lambda }+6 t^2 v+12tuv+10u^2v\right)-4u x \sqrt{\lambda  }(t+u)\bigg]\notag \\
   &&+su(s+t+u)\left(v \left(-m_H^2+s+2 t\right)+x \sqrt{\lambda }\right)\bigg].\label{M22}
\end{eqnarray}

\begin{eqnarray}
&&\vert\mathcal{M}^{N\pm}_{loop}\vert^2= \frac{s}{2} \bigg[(\vert \mathcal{C}_1\vert^2+\vert \mathcal{C}_2\vert^2)t^2+(\vert \mathcal{C}_3\vert^3+\vert \mathcal{C}_4\vert^2)u^2+\frac{m_{\ell}^2}{2}\left(\vert \mathcal{C}_1\vert^2(m_{\ell}^2-2t)+\vert \mathcal{C}_3\vert^2(m_{\ell}^2-2u)\right)\bigg]\notag \\
&&+\frac{ m_{\ell}^2}{2}\bigg[\vert \mathcal{C}_2\vert^2 \left(-4 m_{\ell}^4+m_{\ell}^2 (s+8 t)-2 t
   (s+2 t)+\vert \mathcal{C}_4\vert^2 \left(-4 m_{\ell}^4+m_{\ell}^2 (s+8 u)-2 u
   (s+2 u)\right)\right)\bigg]\notag \\
   &&+\frac{1}{2}  m_\ell \bigg[\left(2 m_{\ell}^2-t-u\right) \left(\text{Re}[\mathcal{C}_1\mathcal{C}_3^*](4 m_{\ell}^3-2 m_\ell (t+u))\mp\text{Im}[\mathcal{C}_1\mathcal{C}_3^*](x
   \sqrt{\lambda  sv})\right)\notag \\
   &&+ \left(t-u\right) \left(\text{Re}[\mathcal{C}_2\mathcal{C}_4^*](2 m_\ell (t-u))\mp\text{Im}[\mathcal{C}_2\mathcal{C}_4^*](x
   \sqrt{\lambda  s}v)\right)\bigg].\label{M23}
\end{eqnarray}
\begin{eqnarray}
&&\vert\mathcal{M}^{N\pm}_{tree-loop}\vert^2=\frac{1}{\left(m_{\ell}^2-t\right)
   \left(m_{\ell}^2-u\right)}\bigg[\text{Re}[\mathcal{C}_0\mathcal{C}_1](24 m_{\ell}^7-4 m_{\ell}^5 (s+8 t+6 u)+2 m_{\ell}^3 (t+u) (2
   s+7 t+3 u)\notag \\
   &&-2 m_\ell
   t \left(2 s u+(t+u)^2\right))\pm\text{Im}[\mathcal{C}_0\mathcal{C}_1]\left( m_{\ell}^2 x \sqrt{\lambda sv}
   (s+2 (t+u))- s t x \sqrt{\lambda sv}-4  m_{\ell}^4 x \sqrt{\lambda sv}\right)\notag \\
   &&+\text{Re}[\mathcal{C}_0\mathcal{C}_3](24 m_{\ell}^7-4 m_{\ell}^5 (s+8 u+6 t)+2 m_{\ell}^3 (t+u) (2
   s+7 u+3 t)\notag \\
   &&-2 m_\ell
   u \left(2 s t+(t+u)^2\right))\mp\text{Im}[\mathcal{C}_0\mathcal{C}_3]\left( m_{\ell}^2 x \sqrt{\lambda sv}
   (s+3 t+5u)- (s +t+u)u x \sqrt{\lambda sv}-6  m_{\ell}^4 x \sqrt{\lambda sv}\right)\bigg].\notag \\
   \label{M24}
\end{eqnarray}
}}

\begin{eqnarray}
&&\vert\mathcal{M}^{T\pm}_{loop}\vert^2= \frac{s}{2} \bigg[(\vert \mathcal{C}_1\vert^2+\vert \mathcal{C}_2\vert^2)t^2+(\vert \mathcal{C}_3\vert^3+\vert \mathcal{C}_4\vert^2)u^2+\frac{m_{\ell}^2}{2}\left(\vert \mathcal{C}_1\vert^2(m_{\ell}^2-2t)+\vert \mathcal{C}_3\vert^2(m_{\ell}^2-2u)\right)\bigg]\notag \\
&&+\frac{ m_{\ell}^2}{2}\bigg[\vert \mathcal{C}_2\vert^2 \left(-4 m_{\ell}^4+m_{\ell}^2 (s+8 t)-2 t
   (s+2 t)+\vert \mathcal{C}_4\vert^2 \left(-4 m_{\ell}^4+m_{\ell}^2 (s+8 u)-2 u
   (s+2 u)\right)\right)\bigg]\notag \\
   &&\pm\frac{m_\ell \sqrt{\lambda}x}{2s} \bigg[2\text{Re}[\mathcal{C}_1\mathcal{C}_2^*]\left(4m_{\ell}^4-m_{\ell}^2(s+2(t+u)+st)\right)-\text{Re}[\mathcal{C}_1\mathcal{C}_4^*](4m_{\ell}^2-s)(2m_{\ell}^2-t-u)\notag \\
   &&+\text{Re}[\mathcal{C}_2\mathcal{C}_3^*]s(t-u)\bigg]-m_{\ell}^2(2m_{\ell}^2-t-u)^2\text{Re}[\mathcal{C}_1\mathcal{C}_3^*]+\text{Re}[\mathcal{C}_2\mathcal{C}_4^*]m_{\ell}^2(t-u)^2\label{M25}.
\end{eqnarray}
\begin{eqnarray}
&&\vert\mathcal{M}^{T\pm}_{tree-loop}\vert^2=\frac{2}{\left(m_{\ell}^2-t\right)
   \left(m_{\ell}^2-u\right)}\bigg[\text{Re}[\mathcal{C}_0\mathcal{C}_1](24 m_{\ell}^7-4 m_{\ell}^5 (s+8 t+6 u)+ 2m_{\ell}^3 (t+u) (2
   s+7 t+3 u)\notag \\
   &&-2m_\ell t \left(2 s u+(t+u)^2\right))+\text{Re}[\mathcal{C}_0\mathcal{C}_3](24 m_{\ell}^7-4 m_{\ell}^5 (s+8 u+6 t)+2 m_{\ell}^3 (t+u) (2
   s+7 u+3 t)\notag \\
   &&-2m_\ell u \left(2 s t+(t+u)^2\right))\pm\frac{\sqrt{\lambda}x}{\sqrt{s}}\bigg[\text{Re}[\mathcal{C}_0\mathcal{C}_2](16m_{\ell}^6-4m_{\ell}^4(2s+t+3u)+m_{\ell}^2(s^2+6st+2su-2t^2+2u^2)\notag\\
   &&-s^2t)-\text{Re}[\mathcal{C}_0\mathcal{C}_4](4m_{\ell}^2-s)(6m_{\ell}^4-m_{\ell}^2(s+3t+5u)+u(s+t+u))\bigg]\bigg].\notag \\.
   \label{M26}
\end{eqnarray}
where $v\equiv\sqrt{1-4m_{\ell}^2/s}$ , $u=m_H^2+2m_\ell^2-s-t$ and $\lambda=(m_H^2-s)^2$.
    
\end{appendix}

\bibliographystyle{apsrev4-1}

\bibliography{main.bib}

\end{document}